\documentclass[manuscript]{aastex}
\usepackage{multirow}
\usepackage{color}
\usepackage{ulem}

\slugcomment{}

\shorttitle{Cloud-cloud collision as a trigger of the high-mass star formation; a molecular line study in RCW120}
\shortauthors{K. Torii et al.}

\begin{document}


\title{Cloud-cloud collision as a trigger of the high-mass star formation; a molecular line study in RCW\,120}


\author{K. Torii\altaffilmark{1,2}, K. Hasegawa\altaffilmark{1}, Y. Hattori\altaffilmark{1}, H. Sano\altaffilmark{1}, A. Ohama\altaffilmark{1}, H. Yamamoto\altaffilmark{1}, K. Tachihara\altaffilmark{1}, S. Soga\altaffilmark{1}, S. Shimizu\altaffilmark{1}, T. Okuda\altaffilmark{3}, N. Mizuno\altaffilmark{3}, T. Onishi\altaffilmark{4}, A. Mizuno\altaffilmark{5}, and Y. Fukui\altaffilmark{1}}
\affil{$^1$Department of Physics, Nagoya University, Chikusa-ku, Nagoya, Aichi 464-8601, Japan}
\affil{$^2$Sub-Department of Astrophysics, University of Oxford, Denys Wilkinson Building, Keble Road, Oxford OX1 3RH, UK}
\affil{$^3$National Astronomical Observatory of Japan, Mitaka, Tokyo 181-8588, Japan}
\affil{$^4$Department of Astrophysics, Graduate School of Science, Osaka Prefecture University, 1-1 Gakuen-cho, Nakaku, Sakai, Osaka 599-8531, Japan}
\affil{$^5$Solar-Terrestrial Environment Laboratory, Nagoya University, Chikusa-ku, Nagoya 464-8601, Japan}

\email{torii@a.phys.nagoya-u.ac.jp}

\begin{abstract}
RCW\,120 is a Galactic H{\sc ii} region having a beautiful ring shape bright in infrared.
Our new CO $J$=1--0 and $J$=3--2 observations performed with the NANTEN2, Mopra, and ASTE telescopes have revealed that two molecular clouds with a velocity separation of 20\,km\,s$^{-1}$ are both physically associated with RCW\,120.
The cloud at $-8$\,km\,s$^{-1}$ apparently traces the infrared ring, while the other cloud at $-28$\,km\,s$^{-1}$ is distributed just outside the opening of the infrared ring, interacting with the H{\sc ii} region as supported by high kinetic temperature of the molecular gas and by the complementary distribution with the ionized gas.
A spherically expanding shell driven by the H{\sc ii} region is usually discussed as the origin of the observed ring structure in RCW\,120. 
Our observations, however, indicate no evidence of the expanding motion in the velocity space, being inconsistent with the expanding shell model. 
We here postulate an alternative that, by applying the model introduced by \citet{hab1992}, the exciting O star in RCW\,120 was formed by a collision between the present two clouds at a colliding velocity $\sim$30\,km\,s$^{-1}$.
In the model, the observed infrared ring can be interpreted as the cavity created in the larger cloud by the collision, whose inner surface is illuminated by the strong UV radiation after the birth of the O star.
We discuss that the present cloud-cloud collision scenario explains the observed signatures of RCW\,120, i.e., its ring morphology, coexistence of the two clouds and their large velocity separation, and absence of the expanding motion.

\end{abstract}

\keywords{ISM: clouds --- Radio lines: ISM}

\section{Introduction}
Feedback effects from high-mass stars such as intense ultraviolet (UV) radiation, strong stellar winds, and supernova explosions at the ends of their lives provide significant effects on the evolution of the interstellar medium (ISM) and star formation in galaxies.
Understanding the formation of high-mass stars is therefore one of the most pressing issues in the contemporary astronomy. 
Despite numerous theoretical and observational studies attempting to address this issue in the last few decades, the detailed physical processes
involved are still elusive \citep[e.g.,][]{por2010}. 
For instance the gas mass accretion rate is one of the known difficulties; high-mass stars must be formed within a short time to overcome the feedback effects of the forming stars, requiring huge mass accretion rates of 10$^{-4}$\,--\,10$^{-3}$\,$M_\odot$\,yr$^{-1}$ \citep[e.g.,][]{wol1987,tan2003, kru2009, hos2009, kui2011}.

Recent attempts to investigate the mechanism of the high-mass star formation in the super star clusters (SSCs) Westerlund\,2 and NGC\,3603 have provided an important clue to understand the high-mass star formation \citep{fur2009, oha2010, fuk2014}.
SSCs have remarkably concentrated stars with total mass of $\sim10^4$\,$M_\odot$ within a small volume less than 1\,pc$^{3}$, and their origin was not clarified. 
The authors identified a pair of two giant molecular clouds with $\sim10^{5}$\,$M_\odot$ associated with each SSC, where the two clouds has a large velocity separation of 10\,--\,20\,km\,s$^{-1}$.
The large velocity separations cannot be explained by expanding motion of the gas driven by stellar winds or supernovae. 
They suggested that a collision between the two clouds followed by strong shock compression of the molecular gas led to the formation of the massive clusters in a short time less than 1\,Myr. 
In addition, a molecular line study in the Galactic H{\sc ii} region Trifid Nebula (M20), which is dominated by a single O star has shown that M20 is a mimic of Westerlnd\,2 and NGC\,3603, where a collision between two molecular clouds with masses of 10$^3$\,$M_\odot$ triggered the formation of the O star \citep{tor2011}.
Here again, the velocity separation of the two clouds ($\sim$7\,km\,s$^{-1}$) is much larger than the permitted gravitational binding velocity of the two clouds.
These three cases, Westerlund\,2, NGC\,3603, and M20, indicate that, although more samples and quantitative studies are necessary, cloud-cloud collision has a possibility to be one of the important modes of high-mass star formation for a wide mass range, from a single OB star to SSCs.

\begin{figure}
\epsscale{.6}
\plotone{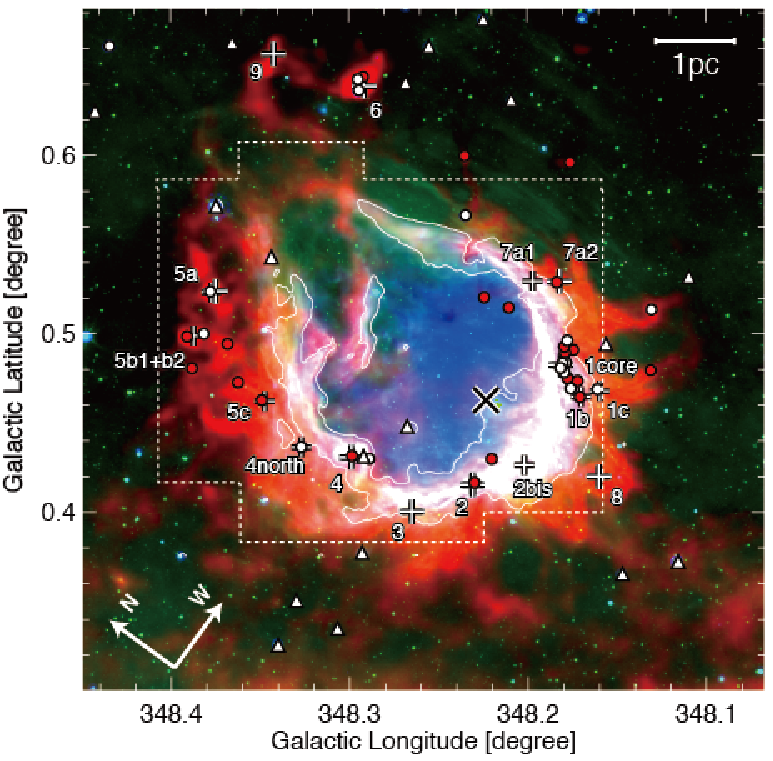}
\caption{A color composite image of RCW\,120. Green, blue, and red show the {\it Spitzer}/IRAC 8\,$\mu$m \citep{ben2003}, the {\it Spitzer}/MIPS 24\,$\mu$m \citep{car2009}, and the {\it Herschel} 250\,$\mu$m \citep{zav2010}. Large cross indicates the exciting star, and filled red circles, filled white circles, and filled white triangles indicate the Class I, intermediate Class I-Class II or flat spectrum, and Class II YSOs identified by \citet{deh2009}. 
Small crosses and labels indicate the cold dust condensations identified at 870\,$\mu$m observations by \citet{deh2009}. Dashed lines show the observed region with Mopra. White contours show the outline of the 8\,$\mu$m ring which is used in the following figures to shows the outline of the 8\,$\mu$m ring, where the {\it Spitzer} 8\,$\mu$m image is median-filtered with a 9$''\times9''$ window.\label{rgb}}
\end{figure}

RCW\,120 is a Galactic H{\sc ii} region \citep{rod1960} located at a distance of 1.3\,kpc from the Sun, having one single exciting O8V or O9V star known as LSS\,3959 or CD$-38^\circ11636$ \citep{rus2003,ave1984, geo1970, zav2007}.  
It shows a beautiful ring-like structure in the mid-infrared and sub-mm wavelengths which encloses the H{\sc ii} region.
Figure \ref{rgb} shows a composite color image of RCW\,120.  
The ring component is clearly traced in the {\it Spitzer}/GLIMPSE 8\,$\mu$m emission \citep[green,][]{ben2003}.
The {\it Spitzer}/MIPSGAL 24\,$\mu$m emission \citep[blue,][]{car2009} is distributed inside of the 8\,$\mu$m ring, which is attributed to hot dust grains illuminated by the UV radiation.
Far infrared emission at 250\,$\mu$m by {\it Herschel}/SPIRE \citep[red,][]{zav2010} also has a ring-like structure, while it is distributed in the slightly outer part of the 8\,$\mu$m ring.
The age of the exciting star in RCW\,120 was investigated by \citet{mar2010}, and the authors discussed that, due to the large uncertainty on the effective temperature and the relative low luminosity of the star, any age less than 5\,Myr is possible.

RCW\,120 is cataloged as the Spitzer bubble S\,7 by \citet{chu2006}. 
They identified about 600 ring-like 8\,$\mu$m structures within $\pm60^\circ$ in the Galactic longitude, and the work was followed by an expanded catalog of 5106 bubbles published by \citet{sim2012}, and many of the bubbles include H{\sc ii} regions \citep{deh2010, ken2012}. 
The pressure-driven expanding H{\sc ii} region \citep[e.g.,][]{wea1977, hos2005} is usually discussed to explain the formation of the Spitzer bubbles \citep[e.g.,][]{deh2010}, in which the expanding ionization front preceded by the shock front driven by the H{\sc ii} region accumulates neutral material between the two fronts, leading formation of a dense layer. 
One can observe this dense layer as an expanding spherical shell on the sky. 
The dense layer is fragmented into gravitationally unstable condensations, and the condensations are expected to form stars \citep[e.g.,][]{hos2005}. 

\citet{zav2007} observed millimeter dust continuum emission in RCW\,120 and found that RCW\,120 is surrounded by dense dust condensations which trace the southern half of the 8\,$\mu$m ring. The condensations were then re-identified with the observations at 870\,$\mu$m by \citet{deh2009} (see small crosses in Figure\,\ref{rgb}).
They estimated the total mass of the ring structure to be about 2000\,$M_\odot$.
\citet{deh2009} also carried out a careful analysis of infrared point sources using the {\it 2MASS} and {\it Spitzer} datasets to identify the young stellar objects (YSOs) in RCW\,120 (see Figure\,\ref{rgb}).
In addition, with the {\it Herschel} observations, \citet{zav2010} discovered one young massive object identified as Class 0 in the most massive condensation (condensation 1core) located at the south-west of the ring, 
The stellar mass estimated for the Class 0 object is 8\,--\,10\,$M_\odot$.
Condensation 1 also harbors a chain of several Class I and Class I-Class II YSOs, which are aligned parallel to the ionization front \citep{zav2007, deh2009}.
These observational signatures indicate that RCW\,120 is a site of on-going star formation.

\citet{zav2007} and \citet{deh2009} discussed the triggered star formation in the ring of the dense neutral medium accumulated by the pressure driven H{\sc ii} region.
They suggested that several different star formation mechanisms are simultaneously working in the ring, e.g., gravitational instabilities, dynamical instabilities, and interaction of the H{\sc ii} region with the pre-existing molecular condensations.
In addition, they discussed that the surrounding ring of RCW\,120 is porous, and the YSOs located distant from the ionization front (e.g., those around the condensation 5 region) could have been formed via interaction with the ionized gas leaking through the ring.

Despite of the detailed studies on the second generation star formation in RCW\,120 in terms of the triggered star formation, the origin of the exciting O star inside the 8\,$\mu$m ring has never been focused on. 
In this paper, we present new molecular observations in CO $J$=1--0 and $J$=3--2 toward RCW\,120 performed with the NANTEN2, Mopra, and ASTE telescopes. 
Mopra and ASTE provide detailed molecular distributions at high angular resolutions, while NANTEN2 is used to describe a large-scale gas distribution, allowing us to have the comprehensive understanding of the origin of RCW\,120.
This paper is organized as follows; Section 2 summarizes the observations and Section 3 the results. 
The discussion is given in Section 4, in which two ideas are tested to understand the observational signatures, i.e., evolution of the H{\sc ii} region and cloud-cloud collision.
Section 5 gives a summary. 

\section{Observations}
The NANTEN2 4-m mm/sub-mm telescope in Atacama, Chile (4,850\,m) was used to image $1.5^\circ \times 1.1^\circ$ of RCW120 in the $^{12}$CO and $^{13}$CO $J$=1--0 emission by using the on-the-fly (OTF) mode during from 2012 May to 2012 December.
A 4\,K cooled SIS mixer receiver provided a typical system temperature of $\sim$250\,K in DSB, and a digital spectrometer provides 16,384 channels at a bandwidth and resolution of 1\,GHz and 61\,kHz, which corresponds to 2600\,km\,s$^{-1}$ with a velocity resolution of 0.17\,km\,s$^{-1}$, respectively, at 110\,GHz. 
We smoothed the obtained data to a velocity resolution of 0.75\,km\,s$^{-1}$ and angular resolution of 200$''$.
The pointing accuracy was confirmed to be better than 15$''$ with the daily observations of the Sun and IRC\,+10216 (R.A., Dec.)(J2000)=($9^{\rm h}~47^{\rm m}~57\fs406$, $13\degr16~\arcmin~43\farcs56$). 
The absolute intensity calibration was done by daily observations of IRAS 16293-2422 in LDN\,1689 (R.A., Dec.)(J2000)=($16^{\rm h}~32^{\rm m}~23\fs3$, $-24\degr~28\arcmin~39\farcs2$) and Perseus molecular cloud (R.A., Dec.)(J2000)=($3^{\rm h}~29^{\rm m}~19\fs0$, $31\degr24~\arcmin~49\farcs0$). 
We compared the obtained data with the FCRAO CO data set, where the data were spatially smoothed to the same angular resolution with NANTEN2, and estimated the main beam efficiency as $\sim$0.53. 
The typical r.m.s. noise fluctuations in the $^{12}$CO $J$=1--0 and $^{13}$CO $J$=1--0 emission are 0.5\,K and 0.4\,K, respectively.

The 22-m ATNF (Australia Telescope National Facility) Mopra mm telescope in Australia was used to image RCW\,120 in $^{12}$CO $J$=1--0, $^{13}$CO $J$=1--0 and C$^{18}$O $J$=1--0 with a high spatial resolution of 33$''$ during 2012 July. 
We covered $16' \times 12'$ of RCW\,120 in the Galactic coordinate, and the OTF mode was used with a unit field of $4'\times4'$.
The typical system noise temperature, $T_{\mathrm{sys}}$, was 400\,--\,600 K in the SSB. The Mopra backend system ``MOPS" provided 4096 channels across 137.5 MHz in each of the two orthogonal polarizations, and the effective velocity resolution was 0.088\,km\,s$^{-1}$ and the velocity coverage was 360\,km\,s$^{-1}$ at 115\,GHz. 
The spectra were gridded to a 15$''$ spacing, and smoothed to a 45$''$ beam size with a 2D Gaussian function. 
The pointing accuracy was checked every 1 hour to keep within 7$''$ by observations of 86\,GHz SiO masers.
We used Orion-KL (R.A., Dec.)=($-5^{\rm h}35^{\rm m}14\fs5$, $-5\degr22\arcmin29\farcs6$) for the absolute intensity calibration and made comparisons with the peak temperature of 100\,K at Orion-KL given by \citet{lad2005} calibrated with an extended beam efficiency introduced in their paper. 
Consistency of the calibrated intensity was then checked by comparing with the NANTEN2 CO $J$=1--0 dataset.
We smoothed the channels in velocity to 0.6\,km\,s$^{-1}$. 
The typical r.m.s. noise fluctuations in the $^{12}$CO, $^{13}$CO and C$^{18}$O $J$=1--0 emission are 0.4\,K, 0.2\,K and 0.2\,K, respectively.

$^{12}$CO$J=3-2$ line observations of RCW120 were performed during 2014 June by using the ASTE 10-m telescope situated in Atacama, Chile (4850\,m). The SIS mixer receiver ``CAT345'' and the digital spectrometer ``MAC'' which provided 512\,MHz  ($\backsimeq445$\,km\,s$^{-1}$) bandwidth and 0.5\,MHz ($\backsimeq0.43$\,km\,s$^{-1}$) resolution were used.  
The observations were made with the OTF mode at a grid spacing of 7\farcs5 toward the same area with the Mopra observations, and the HPBW was 22$''$ at the CO $J=3-2$  frequency.
The pointing accuracy was checked every 1\,--\,2 hours to be better than 2$''$, and the absolute intensity calibration was made by monitoring W28 (R.A., Dec.)(B1950)=($17^{\rm h}57^{\rm m}26\fs8$, $-24\degr03\arcmin54\farcs0$), W44 (R.A., Dec.)(B1950)=($18^{\rm h}50^{\rm m}46\fs1$, $1\degr11\arcmin11\farcs0$), and IRC10216  (R.A., Dec.)(B1950)=($9^{\rm h}45^{\rm m}14\fs8$, $13\degr30\arcmin40\farcs0$) every 1.5 hours. The obtained spectra were compared with the CSO 10\,m data \citep{wan1994} and then the main beam efficiency was estimated to be $0.62 \pm 0.08$. The obtained data were smoothed to the angular and velocity resolutions same as the Mopra dataset, and the typical r.m.s. noise fluctuations are finally 0.1\,K.

\section{Results}
\subsection{Large scale molecular distributions with NANTEN2}
\begin{figure}
\epsscale{.45}
\plotone{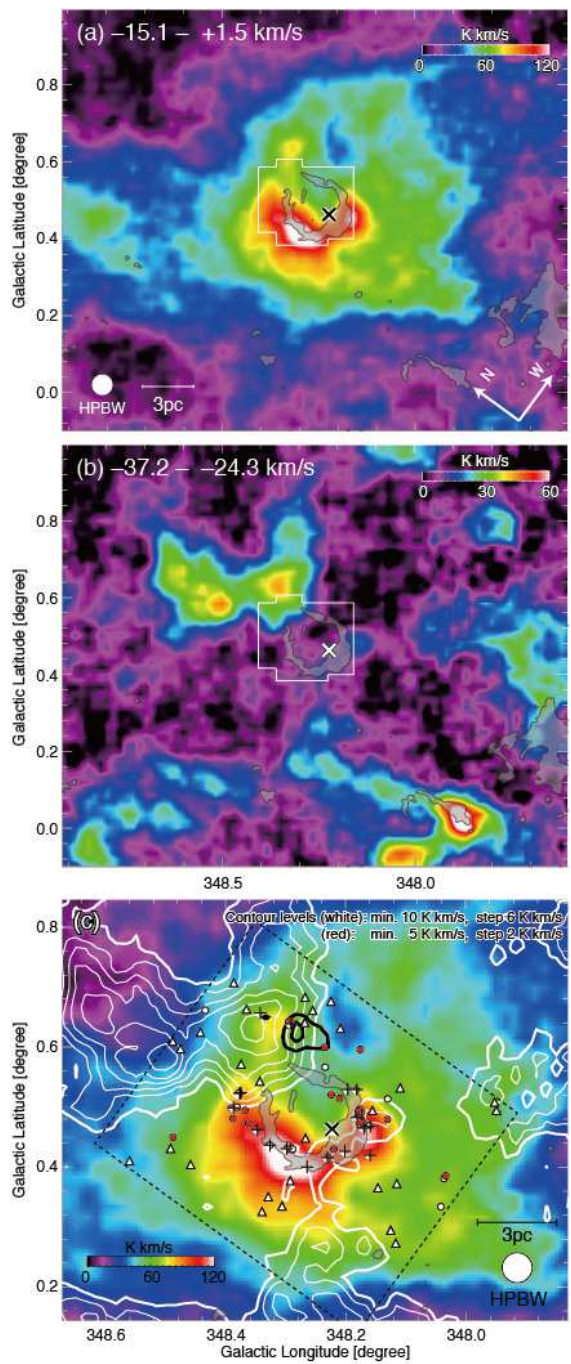}
\caption{Large-scale $^{12}$CO $J$=1--0 distributions towards RCW\,120 are presented with the NANTEN2 dataset. The red cloud and The blue cloud are shown in (a) and (b), respectively. 
The observed area in the Mopra observations is here shown by white lines.
(c) A comparison of the red cloud (image) and the blue cloud (white contours) is shown in a close-up view, in which the bridging feature at $-23$\,--\,$-20$\,km\,s$^{-1}$ shown in Figure\,\ref{nasco_lv} is plotted in the thick black contours.
The YSOs and dust condensations are plotted with the same manner as Figure\,\ref{rgb}, where the region which was used for the YSO identifications is shown by black dashed lines. 
\label{nasco}}
\end{figure}

Using the NANTEN2 $^{12}$CO dataset, molecular gas distributed toward RCW\,120 is identified mainly in two velocity ranges.
Figures \ref{nasco}(a) and (b) show large scale NANTEN2 $^{12}$CO integrated intensity distributions in the two velocity ranges.
In the red-shifted velocity range at $v_{\rm LSR} = -15$\,--\,$+2$\,km\,s$^{-1}$ (Figure \ref{nasco}a), one large molecular cloud which clearly has a ring-like component is distributed at $\sim8$\,km\,s$^{-1}$, which shows an excellent agreement with the 8\,$\mu$m ring, indicating physical association with RCW\,120.
We hereafter refer to the large cloud as ``the red cloud''.
By assuming the flat rotation of the Galactic disk \citep{bra1993}, the distance of the red cloud can be estimated as $\sim$1.3\,kpc with $v_{\rm LSR}$ of $-8$\,km\,s$^{-1}$, which corresponds to the previously known distance of RCW\,120 estimated with the visual extinction towards the central O star \citep{zav2007}.
The size of the red cloud is thus estimated to be about 12\,pc.

In the blue-shifted velocity range at $v_{\rm LSR}$ = $-37$\,--\,$-24$\,km\,s$^{-1}$ shown in Figure \ref{nasco}(b), CO distribution is more complicated and localized relative to that in the red-shifted velocity range. 
A CO cloud having a ``V-shape'' is located at the north of the opening of the 8\,$\mu$m ring at $v_{\rm LSR}$ of $\sim-28$\,km\,s$^{-1}$. 
At each side of the V-shape, one CO peak is seen at its lower part, and the bottom of the right side of the V-shape looks touching the tips of the 8\,$\mu$m ring in projection, suggesting a possible interaction with RCW\,120, although the present coarse spatial resolution cannot establish the physical association with RCW\,120.
We hereafter refer to the V-shape cloud as ``the blue cloud''.
If the blue cloud is associated with RCW\,120, the size of the V-shape is estimated to be about 11\,pc\,$\times$\,7\,pc.
In the blue-shifted velocity range another CO cloud overlapping with the 8\,$\mu$m ring is seen around the southern part of the ring $(l, b\backsimeq 348\fdg20, 0\fdg44)$ at $\sim -33$\,km\,s$^{-1}$ (hereafter ``the $-33$\,km\,s$^{-1}$ cloud'').
It has weak CO intensity and looks connected with the elongated CO features distributed at the southwest of RCW\,120 in  Figure\,\ref{nasco}(b).
The association of the $-33$\,km\,s$^{-1}$ cloud with RCW\,120 is not clear. 

A comparison between the two velocity ranges is shown in a close-up in Figure \ref{nasco}(c), on which the dust condensations and YSOs identified by \citet{deh2009} are superimposed.
The CO peak on the right side of the V-shape looks coinciding with the local CO peak in the red cloud at $(l,b)\backsimeq(348\fdg33, 0\fdg64)$, while the CO peak on the other side seems to have complementary distribution with the rim of the red cloud.
YSOs are extensively distributed throughout the red cloud. 
Class I and Class I-Class II YSOs (red and white circles, respectively) are mainly distributed around the molecular ring component in the red cloud, in which many dust condensations are also concentrated, while Class II YSOs (white triangles) are mainly seen toward the outer part of the red cloud.
Condensations 6 and 9 are distributed toward the coinciding peaks of the blue cloud and the red cloud at the north of the 8\,$\mu$m ring, and several YSOs are also seen there.

\begin{figure}
\epsscale{.6}
\plotone{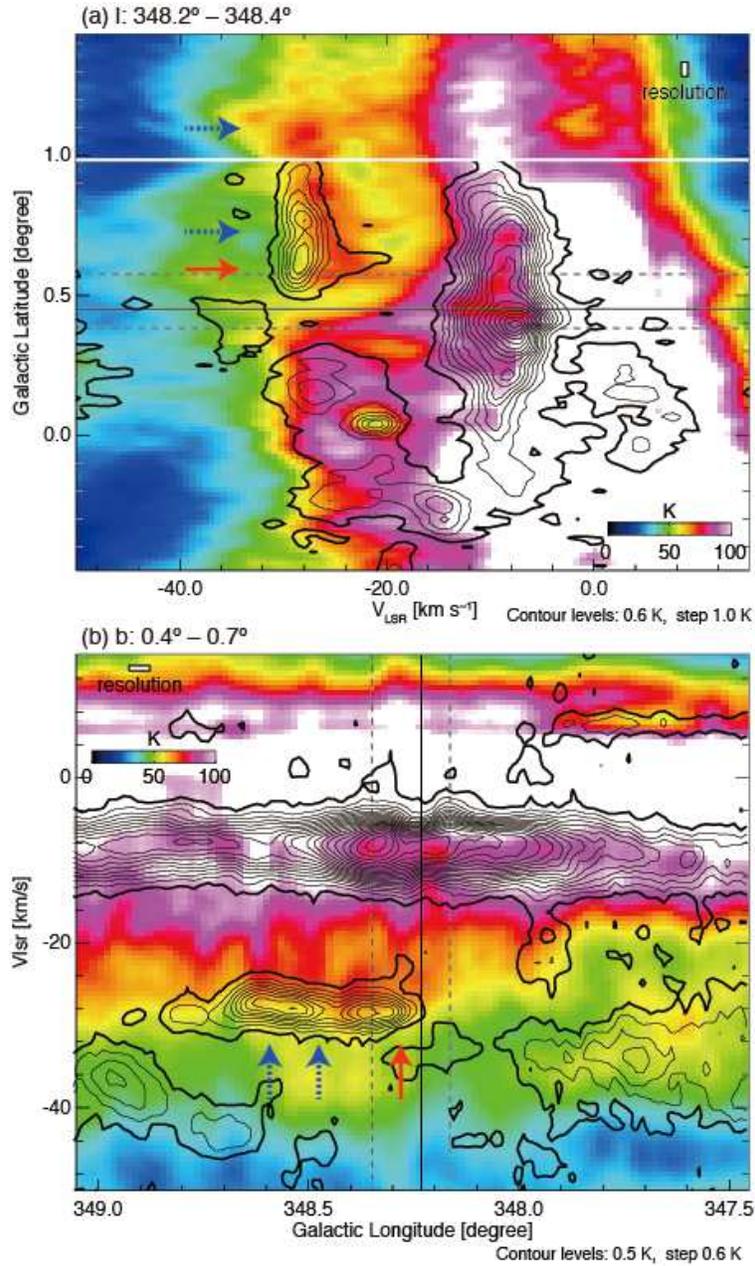}
\caption{Position-velocity maps for a very large area of RCW\,120. Contours show the averaged intensity of the NANTEN2 $^{12}$CO data, and color image shows that of the SGPS H{\sc i} 21-cm data \citep{mcc2005}. Black solid line and dashed lines indicate position of the O star and the approximate extent of the 8\,$\mu$m ring, respectively. The red arrow and blue arrows indicate the directions of the bridging features seen in CO and H{\sc i}, respectively. \label{nasco_lv}}
\end{figure}

Position-velocity diagrams for a large area of RCW\,120 are presented in Figure \ref{nasco_lv} with CO (contours) and H{\sc i} (image), where the SGPS(Southern Galactic Plane Survey) H{\sc i} dataset is used \citep{mcc2005}
As a supplement, CO velocity channel maps including the Mopra datasets are shown in Figures \ref{channel_all0}\,--\,\ref{channel_all4} in Appendix.
The CO emission in the red cloud and the blue cloud have typical velocity widths of $\sim$10\,km\,s$^{-1}$ and $\sim$5\,km\,s$^{-1}$, respectively, and show almost uniform velocity distributions well beyond the extent of RCW\,120.
In Figure\,\ref{nasco_lv} the H{\sc i} emission in the red cloud is seen as an intensity gap in the bright H{\sc i} emission extended over $\pm15$\,km\,s$^{-1}$, while that in the blue cloud is seen at around $-26$\,km\,s$^{-1}$, which is slightly shifted from that in CO around $-28$\,km\,s$^{-1}$.

As depicted by the red arrow in Figure\,\ref{nasco_lv}, the blue cloud has a velocity feature at $(l,b)\backsimeq(348\fdg25, 0\fdg65)$ extended to $\sim-22$\,km\,s$^{-1}$ toward the positive velocity side.
The spatial distribution of the high velocity feature is shown in Figure\,\ref{nasco}(c) with the thick black contour.
It has a size of $\sim0\fdg4$, which corresponds to $\sim1$\,pc at the distance of RCW\,120, and is located just next to the coinciding peaks of the blue cloud and the red cloud.
The corresponding feature is also seen in the H{\sc i} emission (Figure\,\ref{nasco_lv}).
The H{\sc i} feature has a much larger velocity extent and appears to be connecting the blue cloud with the red cloud, suggesting a physical interaction between the two clouds and RCW\,120.
In addition, H{\sc i} has other several high velocity features throughout the blue cloud as shown by the blue arrows with the dashed lines in Figure\,\ref{nasco_lv}.
These H{\sc i} features are not seen at the negative velocity range, and provides another support for the physical connection between the red cloud and the blue cloud even outside RCW\,120.
The typical atomic column density of the high velocity H{\sc i} features is $\sim4\times10^{20}$\,cm$^{-2}$ by assuming optically thin 21-cm emission, and is $\sim8\times10^{20}$\,cm$^{-2}$ by considering the opacity correction of the 21-cm emission discussed in \citet{fuk2014b, fuk2015}.
At a distance of RCW\,120, the total atomic mass of the H{\sc i} features is estimated to be $\sim100$\,$M_\odot$ for the optically thin case, and to be $\sim200$\,$M_\odot$ by adopting the opacity correction by \citet{fuk2014b,fuk2015}.

\subsection{Detailed molecular distributions with Mopra}
\begin{figure}
\epsscale{.95}
\plotone{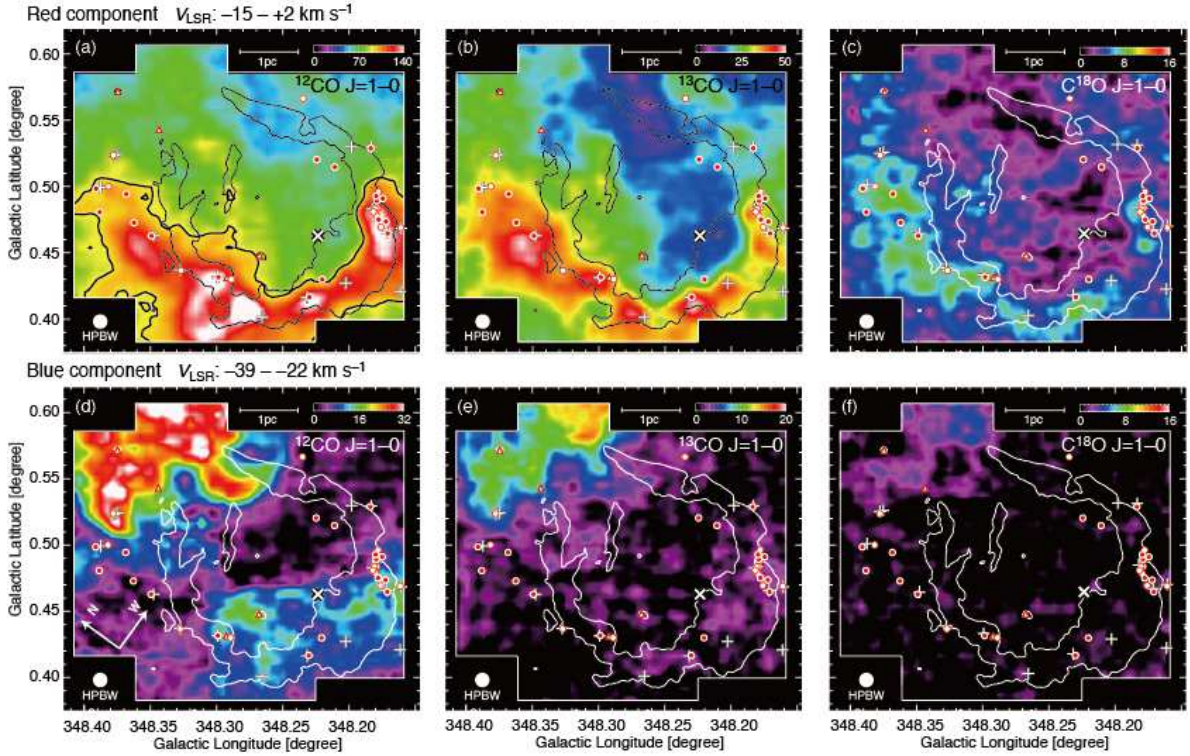}
\caption{$^{12}$CO, $^{13}$CO and C$^{18}$O $J$=1--0 distributions towards RCW\,120 with the Mopra dataset. The red cloud are shown in (a)\,--\,(c), while the blue cloud in (d)\,--\,(f). For the both clouds, $^{12}$CO, $^{13}$CO, C$^{18}$O $J$=1--0 emission are presented in the left, center, right panels, respectively.  The unit of color bars is K km s$^{-1}$. Large cross indicates the O star, and the YSOs and dust condensations are plotted with the same manner as Figure\,\ref{rgb}.  
The gray contours in (a) are plotted at 90\,K\,km\,s$^{-1}$ and indicates the region which is used to estimate the molecular mass of the ring structure (see section 3.5). \label{mopra_lb}}
\end{figure}

Detailed CO distributions toward RCW\,120 using the Mopra datasets are shown at 0.2\,pc spatial resolution.
Figure \ref{mopra_lb} shows the integrated intensity distributions of the $^{12}$CO, $^{13}$CO and C$^{18}$O emission in the red-shifted and blue-shifted velocity ranges.
In Figures\,\ref{mopra_lb}(a)--(c), the red cloud clearly shows a ring-like structure in all the three CO lines, where the molecular ring consists of small clumps with sizes of 0.3\,--\,0.5\,pc.
Detection of C$^{18}$O suggests the existence of the dense gas in the molecular ring, and the dense molecular ring corresponds well with the dust distribution \citep{zav2007,deh2009}.
Typical H$_2$ column density $N$(H$_2$) of the ring component is estimated to be $\sim3\times10^{22}$\,cm$^{-2}$ with the $^{12}$CO integrated intensities by assuming an X-factor of $2\times10^{20}$ (K\,km\,s$^{-1}$)$^{-1}$\,cm$^{-2}$ \citep{str1998}, where the X-factor is an empirical conversion factor from the $^{12}$CO integrated intensity to $N$(H$_2$).
On the other hand, $N$(H$_2$) toward the O star located inside the ring is estimated to be $1.4\times10^{22}$\,cm$^{-2}$.
This estimate can be compared with the visual extinction $A_{\rm v}$ derived as 4.65 mag toward the same direction \citep{ave1984}, which corresponds to the total ISM column density $N$(H{\sc i}+H$_2$) of $9\times10^{21}$\,cm$^{-2}$, almost comparable to the above column densities, by using the relations $N$(H{\sc i}+H$_2$) = $5.8\times10^{21} E(B-V)$\,cm$^{-2}$\,mag$^{-1}$ and $A_{\rm v} = 3.1 E(B-V)$ \citep{boh1978}. 
It is thus suggested that the exciting O star is on the far side of the red cloud along the line-of-sight.

CO distributions in the blue-shifted velocity range obtained with Mopra are shown in Figures\,\ref{mopra_lb}(d)--(f).
The blue cloud is detected in C$^{18}$O with a more than 3\,$\sigma$ level, while the $-33$\,km\,s$^{-1}$ cloud is seen only in $^{12}$CO at the southern part of the 8\,$\mu$m ring.
It is notable that the blue cloud has a molecular protrusion just next to the western tip of the 8\,$\mu$m ring, and it looks elongated inside the 8\,$\mu$m ring through the opening at the north. 
Typical $N$(H$_2$) of the $-33$\,km\,s$^{-1}$ cloud and the blue cloud are estimated with $^{12}$CO to be $\sim4\times10^{21}$\,cm$^{-2}$ and $\sim8\times10^{21}$\,cm$^{-2}$, respectively, significantly smaller than that of the ring component of the red cloud.

\begin{figure}
\epsscale{1.0}
\plotone{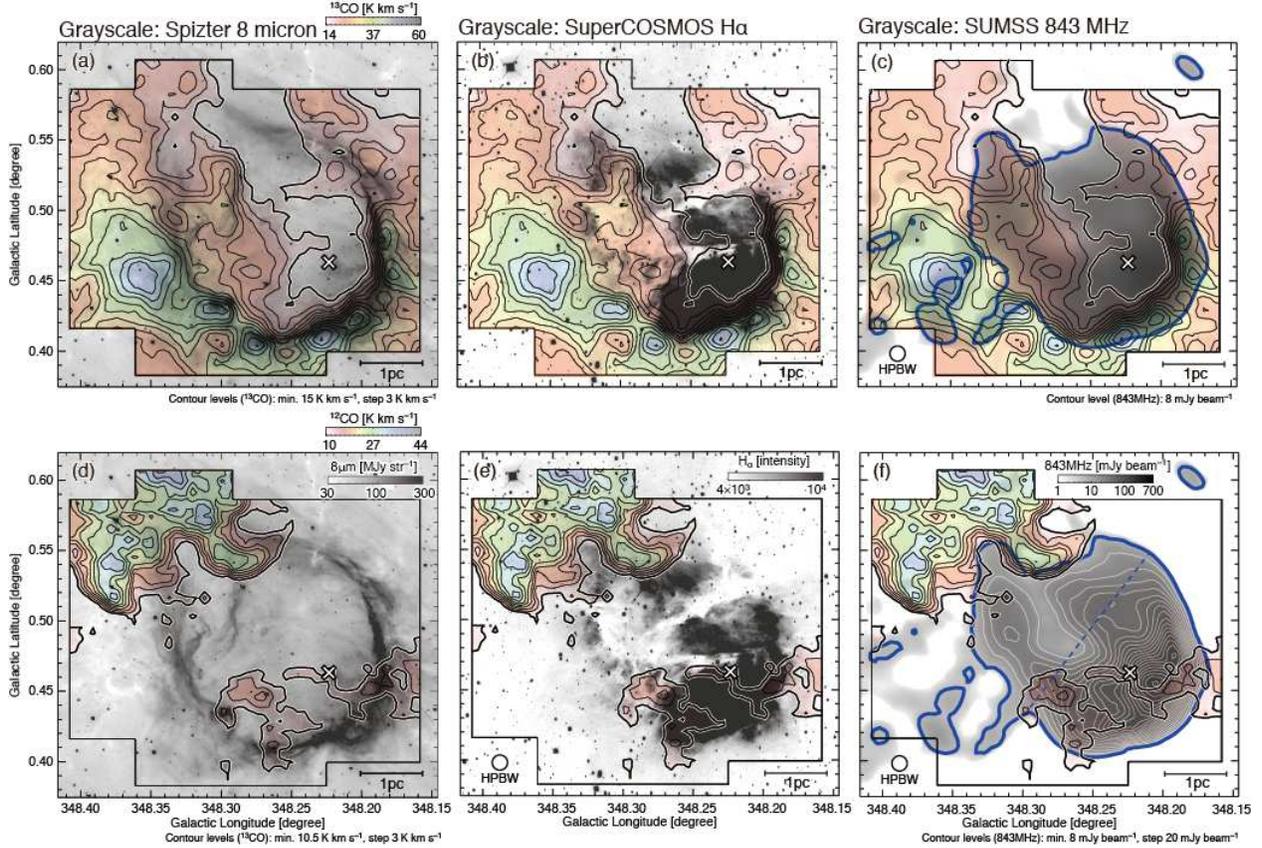}
\caption{(a--c) Comparisons of $^{13}$CO distributions (color and contours) of the red cloud with (a) the {\it Spitzer} 8\,$\mu$m emission (grayscale), (b) the SuperCOSMOS H$\alpha$ (grayscale), and (c) the 843\,MHz SUMSS radio continuum emission (grayscale and thick blue contours). The integration range in velocity is from $-12$\,km\,s$^{-1}$ to $0$\,km\,s$^{-1}$.
(d--f) Comparisons of  $^{12}$CO distributions (color and contours) of the north component of the blue cloud in the same manner as (a)--(c). The integration range in velocity is from $-39$\,km\,s$^{-1}$ to $-22$\,km\,s$^{-1}$. \label{comp}}
\end{figure}

In Figure\,\ref{comp} comparisons between CO and other wavelengths are presented.
Figures\,\ref{comp}(a)--(c) show the $^{13}$CO integrated intensity distributions of the red cloud (black contours and colors) superimposed on (a) the {\it Spitzer} 8\,$\mu$m, (b) the SuperCOSMOS H$\alpha$ \citep{par2005}, and (c) the 843\,MHz radio continuum \citep{boc1999} grayscale images.
In the SUMSS image in Figure\,\ref{comp}(c), the thick contours are plotted at the 5\,$\sigma$ level.
As seen in Figure \ref{comp}(a), the 8\,$\mu$m ring apparently traces the inner rim of the red cloud, supporting physical association between the red cloud and RCW\,120.
No apparent molecular counterpart is seen toward the western tip of the 8\,$\mu$m ring, which is also confirmed with the {\it Herschel} 250\,$\mu$m emission in Figure\,\ref{rgb}.
In the supplemental velocity channel maps in Figure\,\ref{channel_all3}, a possible counterpart for the western tip is seen in $^{12}$CO at around $-10$\,km\,s$^{-1}$, although it is weak in $^{13}$CO and not detected in C$^{18}$O.
Distribution of the ionized gas traced with H$\alpha$ in (b) and the radio continuum emission in (c) is enclosed by the 8\,$\mu$m ring.
The filamentary dark lanes running across the H$\alpha$ image are due to extinction and has CO counterparts in the red cloud, indicating that these molecular counterparts are located on the near side of the H{\sc ii} region.

In Figures \ref{comp}(d)--(f) $^{12}$CO distribution of the blue cloud is used to make comparisons as done for the red cloud in Figures\,\ref{comp}(a)--(c).
As also shown in Figure\,\ref{mopra_lb}(d), a molecular protrusion is elongated from the blue cloud toward inside the 8\,$\mu$m ring through its opening, while the $-33$\,km\,s$^{-1}$ cloud having very weak intensities shows no clear correlation with the infrared emission.
Although the spatial correlation between the blue cloud and the 8\,$\mu$m emission is not apparently identified, the ionized gas clearly shows complementary distribution with the blue cloud, especially with the molecular protrusion, as seen in Figures\,\ref{comp}(e) and (f), where the lowest contours are plotted at the 5\,$\sigma$ level in the radio continuum image.
Since the complementary distribution is seen not only in H$\alpha$ but also in the radio continuum, it is not due to the extinction but is attributed to exclusion of the ionized gas by the blue cloud.
This result suggests that the blue cloud as well as the red cloud is possibly associated with RCW\,120.

\subsection{CO $J$=3--2/$J$=1--0 intensity ratios and LVG analysis}
\begin{figure}
\epsscale{.9}
\plotone{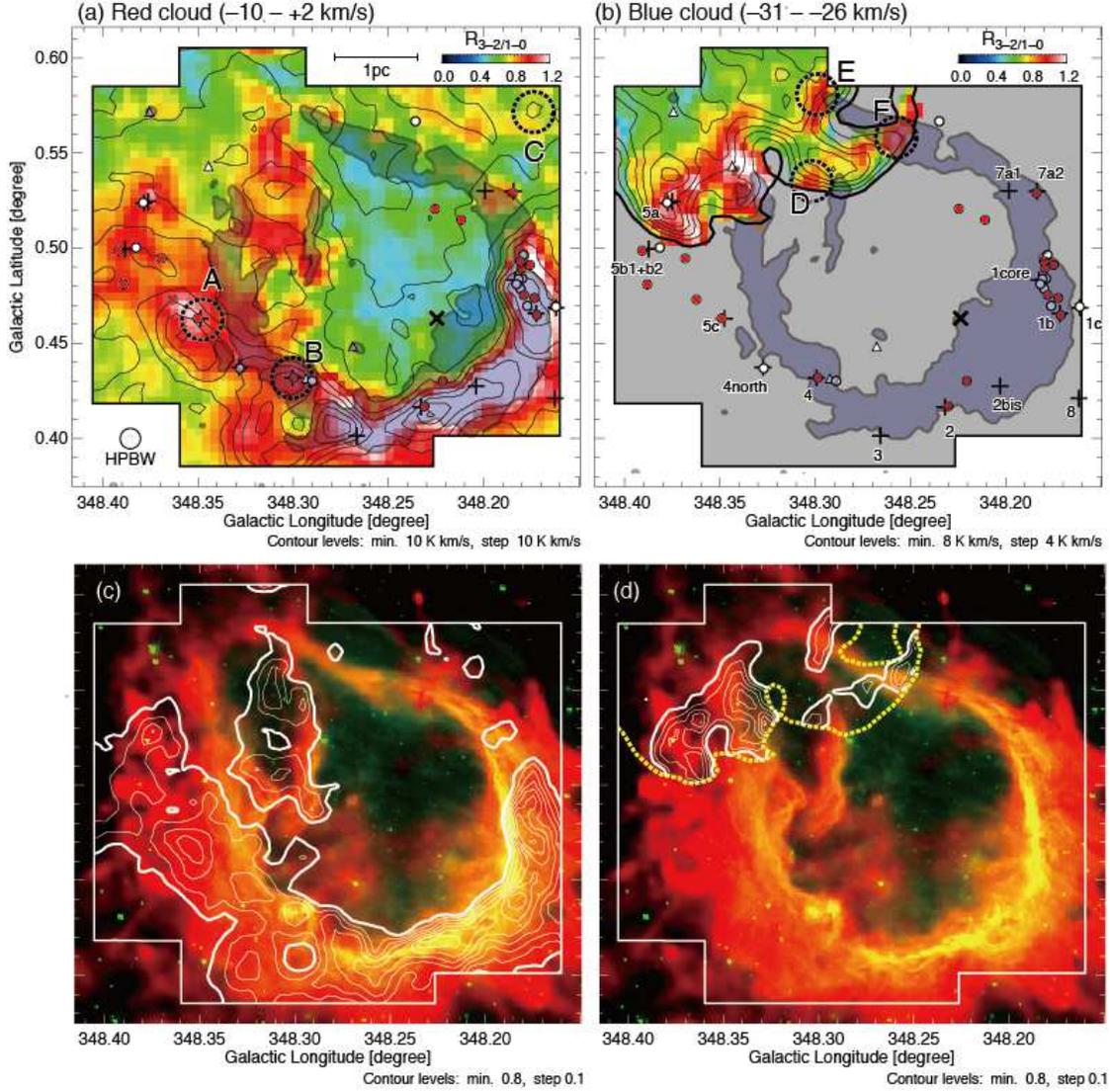}
\caption{
$R_{3-2/1-0}$ maps of the red cloud (a) and the blue cloud (b) are shown in the color image with the $^{12}$CO $J$=1--0 contours. The exciting stars, YSOs, and dust condensations are plotted with the same manner as in Figure\,\ref{rgb}. The names of the condensations given by \citet{deh2009} are shown in (b). Regions A\,--\,F depicted with dashed circles are the regions that are used for the LVG analysis in Figure\,\ref{lvg}.
(c, d) The region with high $R_{3-2/1-0} \geq 0.8$ are shown in contours. The background is a composite image of the {\it Herschel} 250\,$\mu$m (red) and the {\it Spitzer}/IRAC 8\,$\mu$m (green). In (d), the outer boundary of the blue cloud is shown in yellow dashed contours. \label{ratio}
}
\end{figure}

In order to investigate the association of the molecular gas with RCW\,120 into more detail, distributions of the $^{12}$CO $J$=3--2/$J$=1--0 intensity ratio (hereafter $R_{3-2/1-0}$) are presented with the Mopra and ASTE datasets. 
Intensity ratios between different $J$ transitions of CO is useful to diagnose association of the molecular gas with H{\sc ii} regions \citep[see e.g.,][]{tor2011,fuk2014}.
$R_{3-2/1-0}$ distributions of the red cloud and the blue cloud are shown in the upper two panels in Figure\,\ref{ratio}, while in the lower two panels the region with high $R_{3-2/1-0}\geq0.8$ is presented in contours superimposed on a composite image of the {\it Herschel} 250\,$\mu$m (red) and the {\it Spitzer} 8\,$\mu$m (green).
$R_{3-2/1-0}$ typically ranges from 0.6 to $\gtrsim1.2$ in both of the two clouds.
Considering that typical $R_{3-2/1-0}$ in the local clouds without any star formation have $R_{3-2/1-0}$ of $\sim0.4$ \citep{oka2007}, $R_{3-2/1-0}$ in the present two clouds is significantly high.

In the red cloud, high $R_{3-2/1-0}$ of $\gtrsim 0.8$ is mainly seen where the dense molecular ring is located, which is also traced with the 250\,$\mu$m emission (Figure\,\ref{ratio}c). 
The highest $R_{3-2/1-0}$ of $\gtrsim1.2$ is seen around the southern part of the ring, in which condensations 1\,--\,3 are distributed.
Another high $R_{3-2/1-0}$ region is seen inside the 8\,$\mu$m ring at $(l,b)\sim(348\fdg30, 0\fdg50$\,--$0\fdg55)$, whose western rim is traced by the 8\,$\mu$m and 250\,$\mu$ emission, suggesting that the cloud is illuminated by the O star. 
The western tip of the 8\,$\mu$m ring in contrast shows relatively low $R_{3-2/1-0}$ compared with the other part of the 8\,$\mu$m ring.
On the other hand, $R_{3-2/1-0}$ in the blue cloud is high at its southern part. 
The highest $R_{3-2/1-0}$ is seen at an extension of the eastern tip of the 8\,$\mu$m ring, while the western tip also shows high $R_{3-2/1-0}$.
As seen in Figure\,\ref{ratio}(d), these high $R_{3-2/1-0}$ regions show excellent coincidence with the 250\,$\mu$m emission, particularly for the very tip of the western side of the 8\,$\mu$m ring.
Another high $R_{3-2/1-0}$ is distributed toward condensation 5a at the south-eastern rim of the blue cloud, suggesting possible association between the blue cloud and the condensation, although the red cloud also shows high $R_{3-2/1-0}$ toward the same direction as shown in Figure\,\ref{ratio}(a).

In order to investigate the origin of the highly excited gas in both the red cloud and the blue cloud, we here perform the large velocity gradient (LVG) analysis \citep[e.g.,][]{gol1974} to estimate the kinetic temperature $T_{\rm k}$ and number density $n$(H$_2$) of the molecular gas, using the intensity ratios $^{13}$CO $J$=1--0/$^{12}$CO $J$=1--0 (hereafter $R_{13/12}$) as well as $R_{3-2/1-0}$.
It should be noted that the assumption of the uniform velocity gradient is not always valid in the molecular gas associated with H{\sc ii} regions, but radiative transfer calculations considering a microturbulent cloud interacting with an H{\sc ii} region shows no significant difference from the LVG analysis\citep[e.g., ][]{leu1976,whi1977}.
We therefore adopt the LVG approximation in the present study.

Six target regions (regions A\,--\,F) used in the present analysis are shown in Figures\,\ref{ratio}(a) and (b).
Since $^{12}$CO is strongly affected by self-absorption toward the dense part of the ring, particularly around condensations 1\,--\,3, these regions are not adopted.
In the red cloud regions A and B are distributed toward condensations 4 and 5c in the dense ring, respectively, while region C is chosen outside the dense ring to make comparisons with regions A and B.
In the blue cloud all the three regions are chosen toward the high $R_{3-2/1-0}$ regions around the opening of the 8\,$\mu$m ring.

Details of the setup of the analysis is as follows: 
we adopt an abundance ratio $X({\rm CO})$ = [$^{12}$CO]/[H$_2$] = $10^{-4}$ \citep[e.g.,][]{fre1982, leu1984} and [$^{12}$CO]/[$^{13}$CO] = 77 \citep{wilrd1994}, and assume uniform velocity gradient $dv/dr$ in each of the red cloud and the blue cloud as 6\,km\,s$^{-1}$/0.3\,pc $\sim16.7$\,km\,s$^{-1}$\,pc$^{-1}$ and 3\,km\,s$^{-1}$/0.3\,pc $\sim6.7$\,km\,s$^{-1}$\,pc$^{-1}$, respectively.
Errors in $R_{3-2/1-0}$ and $R_{13/12}$ are calculated with 1\,$\sigma$ baseline fluctuations of the spectra and 10\,\% relative calibration error between two CO transitions.
The 10\,\% relative calibration error is applied only to $R_{3-2/1-0}$. 
Since $^{12}$CO $J$=1--0 and $^{13}$CO $J$=1--0 were obtained simultaneously with the same receiver and the same backend system at Mopra, the relative uncertainties between these two transitions are expected to be negligibly small.

\begin{figure}
\epsscale{.85}
\plotone{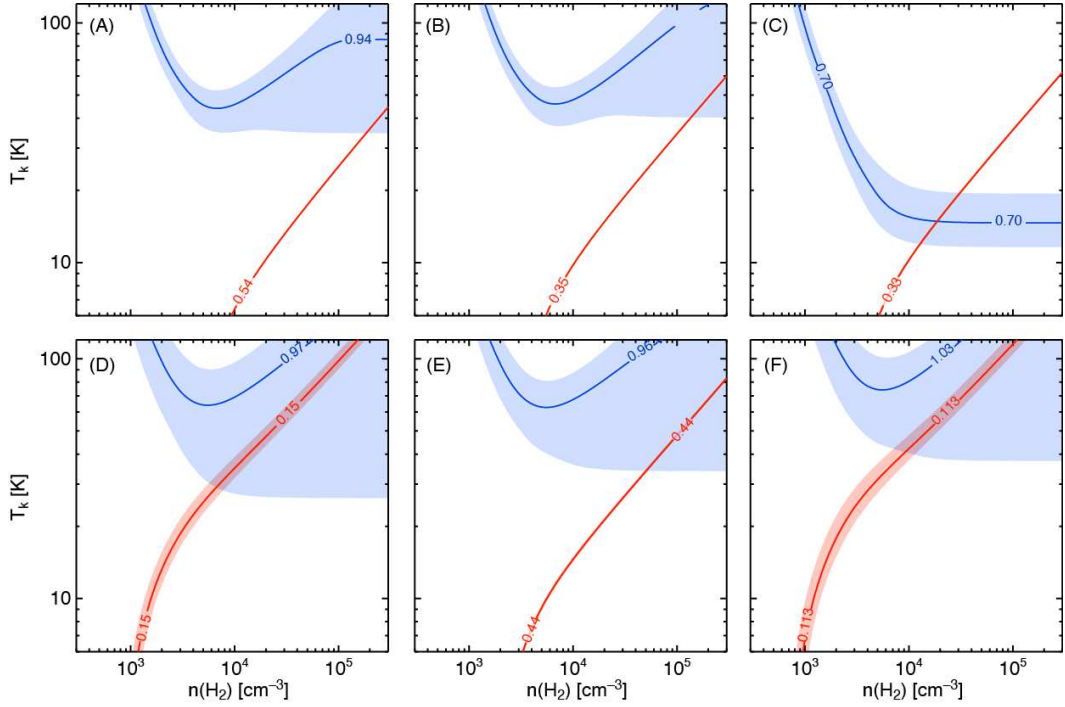}
\caption{
LVG results for the regions A--F in Figure\,\ref{ratio} are shown. $R_{3-2/1-0}$ and its error are shown with blue solid lines and filled blue area, while those of $R_{13/12}$ are shown in red. \label{lvg}
}
\end{figure}

Figure\,\ref{lvg} shows the results of the LVG analysis, in which $T_{\rm k}$ and $n$(H$_2$) are determined where the $R_{3-2/1-0}$ area (blue) and the $R_{13/12}$ area (red) overlap.
Regions A and B at the ring component show high temperature of $\gtrsim40$\,K at high density $\gtrsim10^5$\,cm$^{-3}$, while region C outside the ring indicate relatively low temperature of $\sim 20$\,K at $10^4$\,cm$^{-3}$.
The blue cloud also shows high temperature similar to those in regions A and B, but seems to have lower $n$(H$_2$) of $\gtrsim(1$\,--\,$4)\times10^4$\,cm$^{-3}$.

Possible heating mechanism includes the following two; one is heating by embedded stars in the molecular gas.
The high $T_{\rm k}$ gas with counterparts in YSOs or dust condensations may be understood with this heating, which may be the case in regions A, B, and D.
The other is heating by the feedbacks from the exciting O star, i.e., UV radiation and/or interaction with the H{\sc ii} region.
This heating probably works in the blue cloud, as shown with a negative $R_{3-2/1-0}$ gradient from the south to the north seen in the blue cloud (Figure\,\ref{ratio}b), suggesting that it depends on the distance from the H{\sc ii} region in RCW\,120.
The high $R_{3-2/1-0}$ region which seems to be in contact with the H{\sc ii} region (see Figure\,\ref{comp}) may be interpreted as due to the interaction with the H{\sc ii} region.
On the other hand, the region around the 8\,$\mu$m PAH ring may be understood with the radiative heating by the O star.
Region E may be a case of the interaction with the H{\sc ii} region, while region F keeps both possibilities, and the high $R_{3-2/1-0}$ region at the western tip of the 8\,$\mu$m ring in the blue cloud may be explained with the radiative heating by the O star.

As a summary, the physical association of the red cloud and the blue cloud in RCW\,120 is strongly indicated by the following three results;
\begin{enumerate}
\item High $R_{3-2/1-0}$ and high kinetic temperature ($\gtrsim$30\,--\,40\,K) of the southern part of the blue cloud (Figures\,\ref{ratio} and \ref{lvg}), which coincides with the 250\,$\mu$m image, indicate the heating by the feedbacks of the exciting O star in RCW\,120.
\item The complementary distribution between the blue cloud and the ionized gas tracers (i.e., H$\alpha$ and 845\,MHz emission) supports physical interaction between them (Figure\,\ref{comp}).
\item The intermediate velocity features which connect the red cloud and the blue cloud are seen especially in the H{\sc i} emission (Figure\,\ref{nasco_lv}).
\end{enumerate}

\subsection{Mass of the molecular gas in RCW\,120}
By adopting the same distance of 1.3\,kpc both for the red cloud and the blue cloud, their molecular masses can be estimated by using the integrated intensity of the $^{12}$CO emission with an X-factor of $2\times10^{20}$ (K km s$^{-1}$)$^{-1}$ cm$^{-2}$ \citep{str1998}.
The molecular mass of the entire red cloud shown in Figure \ref{nasco}(a) is calculated to be $5.1\times10^4$\,$M_\odot$ with the NANTEN2 $^{12}$CO dataset, and that of the ring component is to be $4.2\times10^3$\,$M_\odot$ with the Mopra dataset, where the target region is shown by thick black contours plotted at 100\,K\,km\,s$^{-1}$ in Figure \ref{mopra_lb}(a).
On the other hand, the molecular mass of the blue cloud in Figure \ref{nasco}(b) is calculated to be $7.5\times10^3$\,$M_\odot$.

\subsection{Physical condition of the ionized gas in RCW\,120}
To estimate the mass of the ionized gas by measuring the electron density $n_{\rm e}$, the average flux densities of the 843\,MHz emission (Figures\,\ref{comp}c and f) is here measured for the southern half and the northern half of the H{\sc ii} region as separated by the dashed line in Figure\,\ref{comp}(f).
By assuming uniform electron temperature $T_{\rm e}$ of 8000\,K and the path length $L$ of the H{\sc ii} region of 3.5\,pc, the equations 2\,--\,4 in \citet{woo1989} enable us to calculate $n_{\rm e}$ to be 110\,cm$^{-3}$ and 60\,cm$^{-3}$ for the southern half and northern half of the H{\sc ii} region, respectively.
If $T_{\rm e}$ is varying from 5000\,K to 10000\,K, the results change only by $\pm$10\,\%.
The total mass of the ionized gas including ions is thus estimated to be $\sim140$\,$M_\odot$ by assuming a spherical shape of the H{\sc ii} region.

The derived $n_{\rm e}$ is used to estimate the pressure of the H{\sc ii} region as $P_{\rm e} = 2n_{\rm e}\times T_{\rm e} = 1.8\times10^6$\,K\,cm$^{-3}$ for the southern region and $10^6$\,K\,cm$^{-3}$ for the northern region.
These figures can be compared with the turbulent pressure of the surrounding cold neutral medium $P_{\rm n}$. 
The molecular number densities $n$(H$_2$) is given in the LVG analysis (Figure\,\ref{lvg}) as $10^5$\,cm$^{-3}$ for the red cloud and $10^4$\,cm$^{-3}$ for the blue cloud. 
On the other hand, by using the linewidths of the $^{13}$CO profiles, effective temperature of the turbulent gas can be estimated to be 550\,K and 140\,K for the red cloud and the blue cloud, respectively. 
$P_{\rm n}$ in the red cloud is finally given as $5.5\times10^7$\,K\,cm$^{-3}$, one order of magnitude larger than $P_{\rm e}$, while $1.4\times10^6$\,K\,cm$^{-3}$ in the blue cloud, which is nearly equal to $P_{\rm e}$.

These estimates indicate that the H{\sc ii} region cannot drive the bulk motion of the red cloud, while the effect may be marginal for the blue cloud.
It is, however, noted that, as discussed in \citet{wal2011,wal2012}, the ionized gas can stream through less dense gas if the cloud has an inhomogeneous density distributions, and this may be the case in RCW\,120.
As shown in Figures\,\ref{comp}e and f, the distribution of the ionized gas coincides with a gap between the two dense clumps in the blue cloud, implying erosion by the H{\sc ii} region.
On the other hand, the dark 8\,$\mu$m feature seen in north-east of the 8\,$\mu$m ring (Figure\,\ref{comp}) has many filamentary substructures which are radially elongated. For these structures, \citet{zav2007} and \cite{deh2009} discussed leaks of the ionized gas through the dense ring component. 
The leakage of the ionized gas may lead to underestimate of the total mass ionized by the O star.
However, as discussed by \citet{and2014}, the mass fraction of the escaping ionized gas may be minor, about 25\,\% of that observed.

\section{Discussion}

Our new CO observations with NANTEN2 and Mopra have shown that the two CO clouds, the red cloud and the blue cloud, at velocities of $\sim-28$\,km\,s$^{-1}$ and $\sim-8$\,km\,s$^{-1}$ are both associated with RCW\,120.
The red cloud with a mass of $5 \times 10^4$\,$M_\odot$ clearly traces the 8\,$\mu$m ring, while the blue cloud with a mass $\sim8\times10^3$\,$M_\odot$ located just north of the opening of the 8\,$\mu$m ring shows complementary distribution with the H{\sc ii} region.
Both of the two clouds have high kinetic temperature $\gtrsim20$\,--\,40\,K, supporting the coexistence of the two clouds at RCW\,120.

To understand the coexistence of the two clouds, interpreting the large velocity separation of $\sim$20\,km\,s$^{-1}$ may be the key. 
By tentatively assuming an viewing angle of the clouds' relative motion to the line-of-sight as 45$^\circ$, the relative velocity between the two clouds is estimated to be $\sim$30\,km\,s$^{-1}$, which requires the total mass of $\sim2\times10^6$\,$M_\odot$ to gravitationally bind the two clouds within 12\,pc of the exciting star. 
This figure is significantly larger than the total molecular mass in RCW\,120 derived as $\sim6\times10^4$\,$M_\odot$, indicating that the coexistence of the two clouds cannot be understood as a gravitationally bound system.
Even if we adopt the projected velocity separation 20\,km\,s$^{-1}$ as the real relative velocity, the cloud motion is not gravitationally bound.

\subsection{Evolution of the H{\sc ii} region as the origin of RCW\,120}
A possible explanation of the velocity separation between the two clouds is the feedback effects of the exciting star, such as the stellar wind, UV radiation, and pressure driven H{\sc ii} region.
Supernova explosions are also an alternative candidate of the driving source, whereas, in RCW\,120 no supernova remnants have been identified to date.
In the following, in order to explain the morphology of RCW\,120 and coexistence of the two clouds, we examine several ideas in terms of the evolutional scenario of the H{\sc ii} region.

\subsubsection{Cases of uniform ambient medium}
We first investigate the evolution of the H{\sc ii} region as the origin of the two clouds in RCW\,120, by testing whether it can explain the ring-like morphology of RCW\,120, velocity distribution of the molecular gas, and the morphology of the two clouds.
A basic idea to explain the ring shape is known as ``Collect \& Collapse (hereafter C\&C)'', which was first proposed by \citet{elm1977}. 
An expanding H{\sc ii} region sweeps up uniform ambient material between the ionization front and the preceding shock front to form a dense molecular layer \citep[e.g.,][]{hos2005, hos2006, dal2007}.
The dense layer fragments via gravitational instability, and the individual fragments may finally form a new generation of stars \citep{whi1994a, whi1994b, elm1994, dal2009}.
This model is usually discussed as the origin of the Spitzer bubbles including RCW\,120 \citep[e.g.,][]{deh2010}.

The molecular masses accumulated within the dense layer by C\&C, $M_{\rm ring}$, is estimated with a equation $4/3\pi r_{\rm ring}^3 \rho_0$, where $r_{\rm ring}$ is the radius of the ring, and $\rho_0$ is the initial density of the ambient medium which corresponds to number density $n_0$.
By assuming $M_{\rm ring}$ of 3100\,$M_\odot$ derived in Section 3.5, $n_0$ is calculated to be $\sim$3000\,cm$^{-3}$, which is consistent with the estimate of \citet{zav2007}.
\citet{hos2005, hos2006} provide a basic 1D hydrodynamical numerical simulations of C\&C, in which the radiative transfer of UV and far-UV radiation is taken into account.
The case of the model S19 in \citet{hos2006}, in which the mass of the exciting star is 19.0\,$M_\odot$, provides nearly the same situation of RCW\,120, and by using their equation (31) and (36), the formation timescale of RCW\,120 with a ring radius of 1.7\,pc can be estimated to be $\sim$0.4\,Myr \citep[see Figures\,14 and 15 in][]{zav2007}.

\begin{figure}
\epsscale{.5}
\plotone{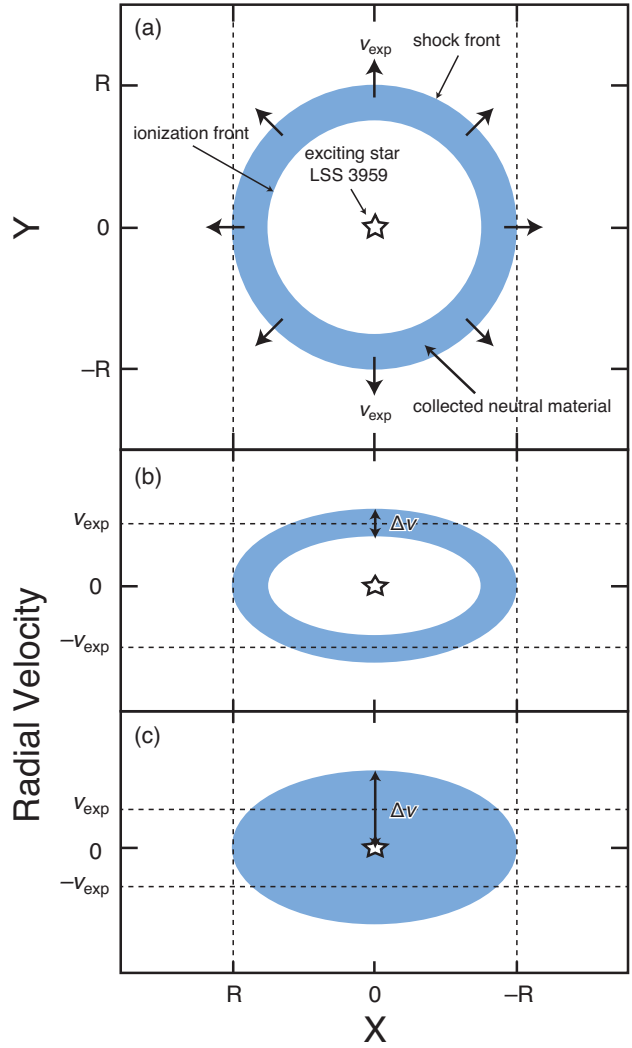}
\caption{Schematic images of the expanding shell model. Spatial distribution is shown in (a), while position-velocity maps expected in the model with different ratios of $v_{\rm exp}$ to $\Delta v$ are shown in (b) and (c). \label{bubble_model}}
\end{figure}

A schematic of the characteristic velocity distribution predicted by C\&C is shown in Figure\,\ref{bubble_model}.
If the dense layer surrounding the H{\sc ii} region expands at an expanding velocity $v_{\rm exp}$ because of high pressure driven by the ionized gas, the expanding spherical layer is expected to have velocity distributions like those shown in Figures \ref{bubble_model}(b)\,--\,(c).
If velocity width, $\Delta v$, is small enough, the near and far sides of the layer can be separately observed toward the exciting star with a velocity separation of 2$v_{\rm exp}$ (Figure\,\ref{bubble_model}b).
On the other hand, if $\Delta v$ is too large compared with $v_{\rm exp}$, the near and far sides cannot be distinguished, and only one single component with an elliptical shape may be observed, where the direction of the O star has the largest velocity width (Figure\,\ref{bubble_model}c).

\begin{figure}
\epsscale{.9}
\plotone{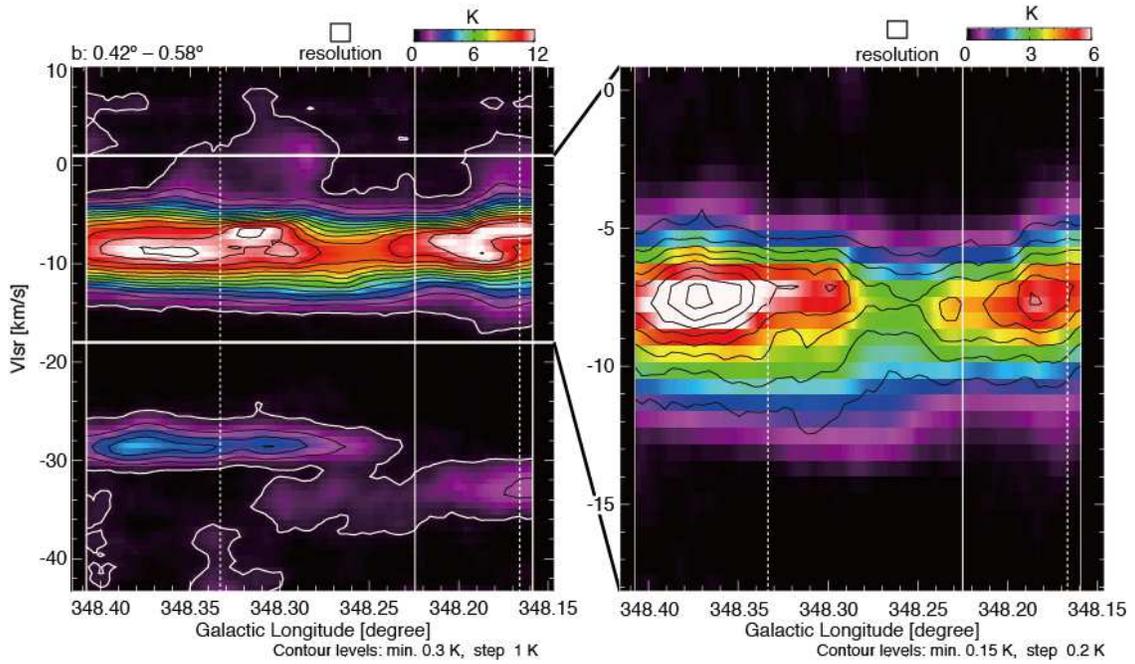}
\caption{(a) Longitude-velocity map of the Mopra $^{12}$CO dataset. Vertical solid line and dotted lines indicate the position of the O star and approximate extent of the 8\,$\mu$m ring. (b) A close up of the red cloud in $^{13}$CO (color) and C$^{18}$O (contours). \label{mopra_lvall}}
\end{figure}

Figure\,\ref{mopra_lvall} shows the longitude-velocity map of the Mopra $^{12}$CO data at a latitude range 0\fdg42\,--\,0\fdg58, which is complementary to that of the NANTEN2 dataset in Figure\,\ref{nasco_lv}.
$v_{\rm exp}$ at the time 0.4\,Myr is estimated to be $\sim$2\,km\,s$^{-1}$ \citep{zav2007}.
We here test two cases; one is the case that both of the red cloud and the blue cloud were formed by C\&C (case I), and the other is that only the red cloud which shows a ring-like shape was formed through C\&C, and the blue cloud is associated with RCW\,120 by some other reason (case II).

For the case I, as clearly seen in Figures\,\ref{nasco_lv} and \ref{mopra_lvall}(a), both of the red cloud and the blue cloud have uniform velocity distributions well beyond the extent of RCW\,120, and furthermore, the velocity separation of $\sim$20\,km\,s$^{-1}$ is much larger than the expected velocity separation of 4\,km\,s$^{-1}$ in the C\&C scenario.
To discuss the case II, $^{13}$CO and C$^{18}$O velocity distributions of the red cloud is shown in Figure \ref{mopra_lvall}(b).
Our datasets indicate only one velocity component toward the exciting star, and the velocity width toward the exciting star is smaller than those in the other parts, showing no signature of the expanding motion.

\begin{figure}
\epsscale{1.}
\plotone{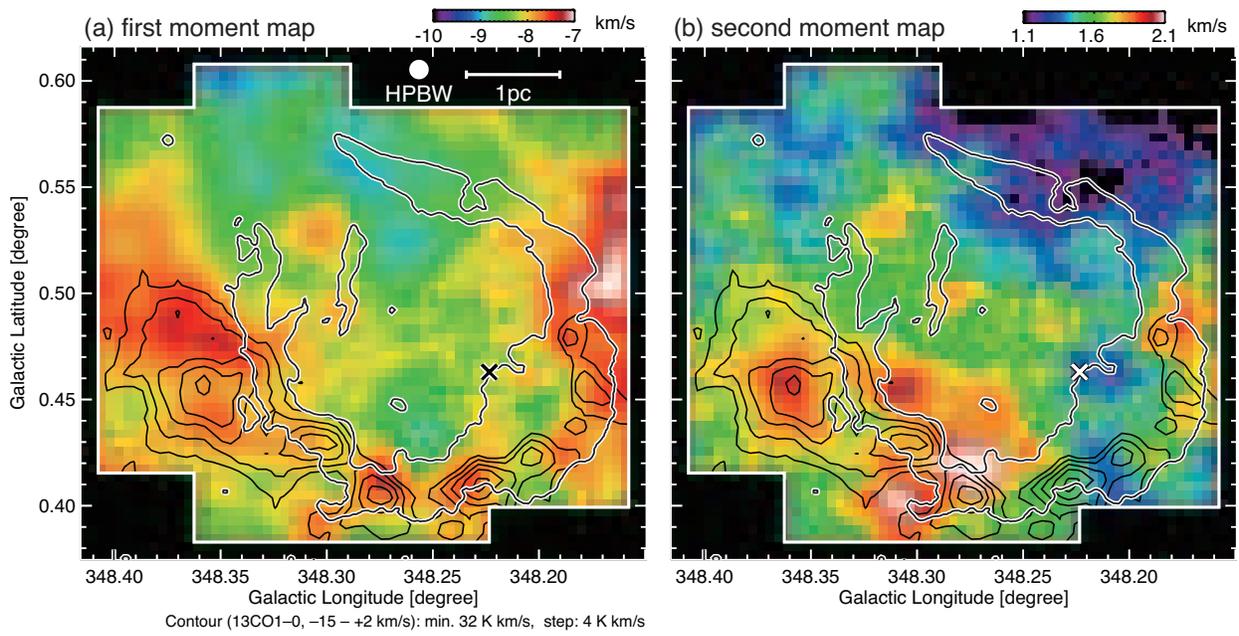}
\caption{(a) The first moment map and (b) the second moment map of $^{13}$CO $J$=1--0 in the red cloud. The contours shows the dense molecular ring in $^{13}$CO $J$=1--0 integrated over $-15$\,--\,$+2$\,km\,s$^{-1}$.
\label{peakv+dv}}
\end{figure}

In order to show the velocity distributions into more detail, intensity-weighted velocity map and velocity dispersion map are shown by the first-moment map and the second-moment map, respectively, in Figures\,\ref{peakv+dv}.
Although velocities are slightly shifted from the ring component ($\sim-8$\,km\,s$^{-1}$) to the ring interior ($\sim-8.5$\,km\,s$^{-1}$) as seen in Figure\,\ref{peakv+dv}(a), the velocity dispersion shown in Figure\,\ref{peakv+dv}(b) is increased toward the ring component ($\sim2$\,--\,$2.5$\,km\,s$^{-1}$) relative to that toward the ring interior and the exciting O star ($\sim1.5$\,km\,s$^{-1}$), which is an opposite trend to what is expected in the C\&C model.

If the H{\sc ii} region is in the pressure equilibrium with the surrounding neutral medium, which is possible as discussed in Section 3.5, the expanding motion may not be observable, and the observational signatures may become consistent with the C\&C model.
\citet{hos2006} provide the relationship between the expanding velocity and the time in their equation 44.
To reduce $v_{\rm exp}$ to be less than our detection limit of 0.5\,km\,s$^{-1}$ at the given $n_0$ of 3000\,cm$^{-3}$ in 5\,Myr, which is the safe upper limit of the age of RCW\,120 \citep{mar2010}, the UV flux of the exciting star is required to be $<10^{45.2}$\,photons\,s$^{-1}$ \citep[see the equation 40 of][]{hos2006}. 
It is two orders of magnitude smaller than that of the exciting star in RCW\,120 \citep[e.g.,][]{zav2007}, and it is therefore difficult to explain the observations with the pressure equilibrium assumption.

\subsubsection{Cases of non-uniform ambient medium}
Recent theoretical studies have investigated evolution of H{\sc ii} regions in non-uniform ambient medium.
\citet{wal2011,wal2012} presented numerical calculations of the evolution of an H{\sc ii} region in a spherical cloud having fractal density structures.
In particular, in \citet{wal2011}, the authors investigated the formation of RCW\,120 and successfully reproduced the morphology, size, and mass of the neutral medium in RCW\,120.
In their calculations, the ionized gas streams out through less dense regions rather than dense regions, and the pressurized low-density region compresses the remaining gas, leading to the formation of dense clumps in the cloud.
The resultant density distribution strongly reflects the initial condition.
In their study, the shell-like structure is still formed by expansion, and the cold shell material is continuously accelerated up to 4\,km\,s$^{-1}$ \citep{wal2012}. 
The discussion for case I and case II with the uniform medium case holds valid; no expanding motion having a velocity separation of 8\,km\,s$^{-1}$ is seen in RCW\,120.

\begin{figure}
\epsscale{1.}
\plotone{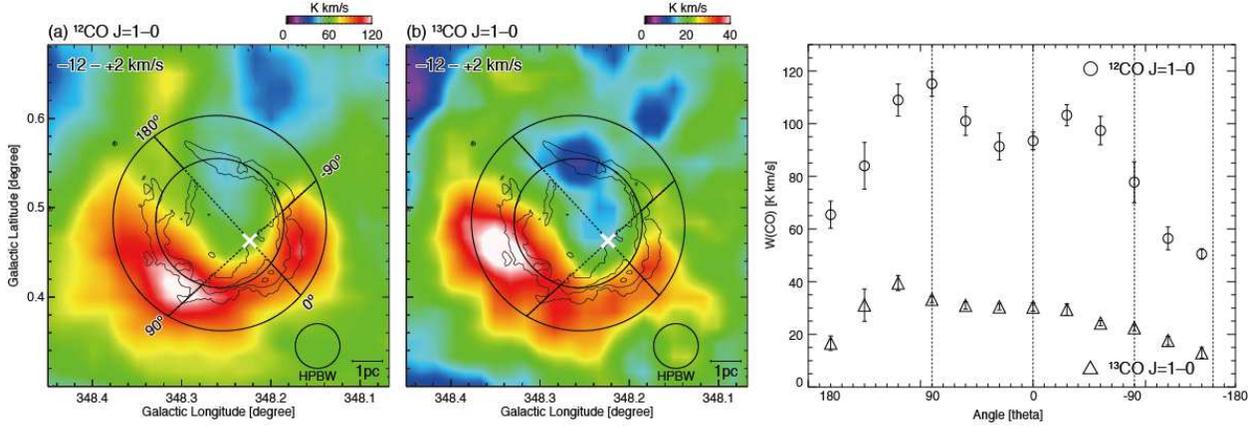}
\caption{Azimuthal distributions of the red cloud with the NANTEN2 $^{12}$CO and $^{13}$CO datasets are shown in (c), where the target regions of the plots are shown in (a) and (b) as a ring superimposed on the $^{12}$CO and $^{13}$CO integrated intensity maps.
In (a) and (b), the inner boundary of the target ring is determined by fit to the 8~$\mu$m image with a elliptical function, and the ring with 1.5 times larger size is then plotted as the outer boundary. The resulting major axes of the inner and outer rings are 1.6\,pc and 2.4\,pc, respectively, and the eccentricity is 0.24.
The center of the ellipse is estimated to be $(l,b)=(348\fdg257, 0\fdg483)$, while the center of the azimuthal plots is taken at the position of the O star.
In the azimuthal plots in (c), Az = 0$^\circ$ is defined at the point where the distance from the O star to the ring is smallest. \label{radial_plot}}
\end{figure}

\citet{och2014} calculated evolution of an H{\sc ii} region by assuming a Bonnor-Ebert sphere, in which the exciting star has an offset from the center of the sphere.
In their calculations, first C\&C works to form a dense layer, but the layer is eventually broken by the over-pressured ionized gas at the thinnest point of the layer. 
The ionized gas then streams into the surrounding ISM, forming the observed morphology of RCW\,120.
This scenario can explain the offset position of the exciting star from the center of the 8\,$\mu$m ring, as well as the morphology of RCW\,120.
The formation timescale they estimated is $\sim$3\,Myrs, also consistent with the observational limit ($<5$\,Myrs).
To make a comparison with their model, azimuthal distributions of the NANTEN2 CO maps are presented in Figure\,\ref{radial_plot}, where two inner and outer ellipses comprise the target ring as shown in Figures\,\ref{radial_plot}(a) and (b). 
The inner ellipse is determined by fit to trace the inner rim of the 8\,$\mu$m ring, and the outer one is plotted to have a 1.5 times larger size to fully cover the strong CO emission.
The position angle of the ellipse is 47$^\circ$ in a counterclockwise direction.
In the scenario presented by \citet{och2014}, CO emission should be strongest at the point where the distance from the exciting star to the ring is smallest, which here corresponds to Az = 0$^\circ$, and should be weakest around the opening of the 8\,$\mu$m ring (Az $\sim$ 180$^\circ$).
The observed CO emission is, however, strongest at Az $\sim$ 100$^\circ$, and weakest at Az $\sim$ $-150^\circ$, which is inconsistent with their scenario.

Here is a brief summary of the above discussion. In the previous studies, interpretations of RCW\,120 based on the evolution of an H{\sc ii} region are basically successful in reproduction of the morphology and mass of RCW\,120. However, the velocity distribution revealed in the present observations shows no expanding motion contrary to what the model predicts. 

\subsection{An alternative: Cloud-Cloud Collision}

We here postulate an alternative idea that an accidental collision between the red cloud and the blue cloud triggered the formation of the O star and the surrounding ring component in RCW\,120. 
The idea provides an explanation for the coexistence of the two clouds having a large velocity separation as well as the other issues laid on this object.

\subsubsection{Previous studies on cloud-cloud collision}

There has been discussion that collisions between the interstellar molecular clouds may trigger the star formation \citep[e.g.,][]{lor1976, hig2010, dua2011}.
Many of the previous studies reported discoveries of pairs of molecular clouds having velocity separations of a few km\,s$^{-1}$.
Since the small velocity separation can be understood not only as Cloud-Cloud Collision (CCC) but also as the gravitationally bound system, these works do not provide compelling evidence for the CCC.
Another possible star forming site induced by CCC was found in the Galactic center \citep{has1994}. 
This is a case having large relative velocity in the collision, $\sim$30\,km\,s$^{-1}$, with triggered formation of O stars and the gravitational binding is unlikely, making the CCC as a plausible interpretation.
This however does not provide a case of CCC-triggered star formation common throughout the Galactic disk, since the physical conditions in the Galactic center are significantly different, including highly turbulent state, from that in the Galactic disk. 
It is only recent that robust evidence for the CCC and its role on the high-mass star formation has been presented observationally in the Galactic disk, even the number of the case is still limited.
In the observational studies on Westerlund\,2, NGC\,3603, and M20, associations of the two molecular clouds having a large velocity separation of $\sim10$\,--\,20\,km\,s$^{-1}$ were shown \citep{fur2009,oha2010,fuk2014, tor2011}.
These velocity separations are too large to be explained by the gravitational motion of a bound orbit, and an accidental collision is strongly suggested.
Recent finding of another high-mass star formation triggered by cloud-cloud collision in the N159W with ALMA lends support for the important role of CCC even in the Large Magellanic Cloud \citep{fuk2015b}.

A pioneering numerical study of CCC was carried out by \citet{hab1992}, followed by \citet{ana2010} and \citet{tak2014}.
These calculations indicate that CCC works as a trigger to form massive molecular clumps whose masses are larger than their Jeans mass.
In addition, \citet{ino2013} presented 3D magnetohydrodynamic calculations of CCC and discussed that the magnetic fields can help to form massive molecular filaments and clumps by driving gas flow to convex points of the deformed shock wave, leading large mass accretion rates of $10^{-4}$\,--\,$10^{-3}$\,$M_\odot$\,yr$^{-1}$.
Such collisions are expected to be frequent in gas-rich galaxies as discussed by \citet{tas2009} and \citet{dob2015}.

\subsubsection{Basic scenario of the Habe \& Ohta model}
In order to understand the present observations in RCW\,120, we first discuss the basic idea of the CCC scenario presented by \citet{hab1992}, in which collisions between two molecular clouds having different sizes were investigated.
Because the authors did not include star formation and subsequent UV radiation, in the schematic shown in Figure \ref{ccc_model} some relevant events are qualitatively added.
First, a small cloud and a large cloud are approaching each other (stage 0).
Once a collision occurs, a dense layer is formed by compression at the interface of the two clouds, creating a cavity in the large cloud (stage 1). 
The diameter of the cavity is nearly equal to that of the small cloud.
The compressed layer is highly turbulent, which can be observed as an intermediate velocity structure \citep{haw2015}, and the small cloud streams into the compressed layer during the collision. 
Along with the cavity creation, massive clump formation and subsequent high-mass star(s) formation occur in the compressed layer, followed by the strong UV radiation from the high-mass star(s) which ionizes the surrounding neutral material (stage 2).
At the final stage (stage 3), the ionized gas fills up the cavity and erodes the inner surface of the cavity which includes the compressed layer, enhancing the PAH emission observable at 8\,$\mu$m.

Stage 3 may have variations depending on the velocity of the compressed layer.
For example, if deceleration is fast, cavity creation almost stops and much molecular gas remains forward of the moving direction (stage 3A).
On the other hand, if the velocity keeps fast, the dense layer completely penetrates the large cloud, forming a tunnel instead of a cavity (stage 3B).
In either case, if the ionized gas completely escapes, only the cavity or the tunnel is observable.
The depth of the cavity and the location of the O star in the cavity are determined by density of the clouds and physical lengths of the small cloud along the colliding direction.
A detailed analysis of the cavity creation and its effects on the star formation will be presented in a forthcoming paper (Torii et al. 2015b, in prep.).

The resultant cavity/tunnel should be observed as a projection on the sky.
This is just like looking at a ``wine glass'' (stage 3A) or a ``tube'' (stage 3B) from various angles.
If observing from the bottom or the top of the glass/tube, one can see a ring shape, and if from the sides, one may see a U-shape or two separated components as shown in Figure\,\ref{ccc_model} as 3A and 3B, respectively.

It should be noted that, in the Habe \& Ohta model it is assumed that two molecular clouds with different sizes collide. 
On the other hand, in M20 the observed colliding clouds have almost the same sizes \citep{tor2011}, and in such a case creation of the cavity is not expected.
It is likely to be more frequent that collisions take place between clouds with different sizes, and the case in M20 is thus relatively rare.

\subsubsection{Cloud-cloud collision in RCW\,120}
\begin{figure}
\epsscale{1.}
\plotone{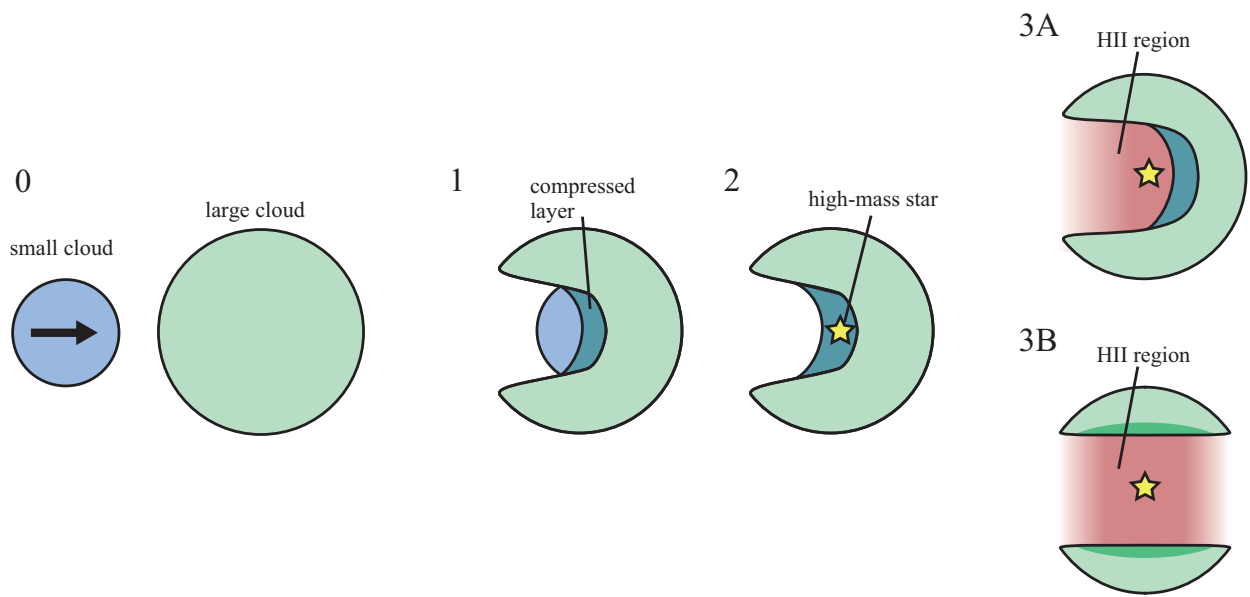}
\caption{Schematics of the basic scenario of CCC introduced by \citet{hab1992} (see section 4.2.2 for details). \label{ccc_model}}
\end{figure}

\begin{figure}
\epsscale{.8}
\plotone{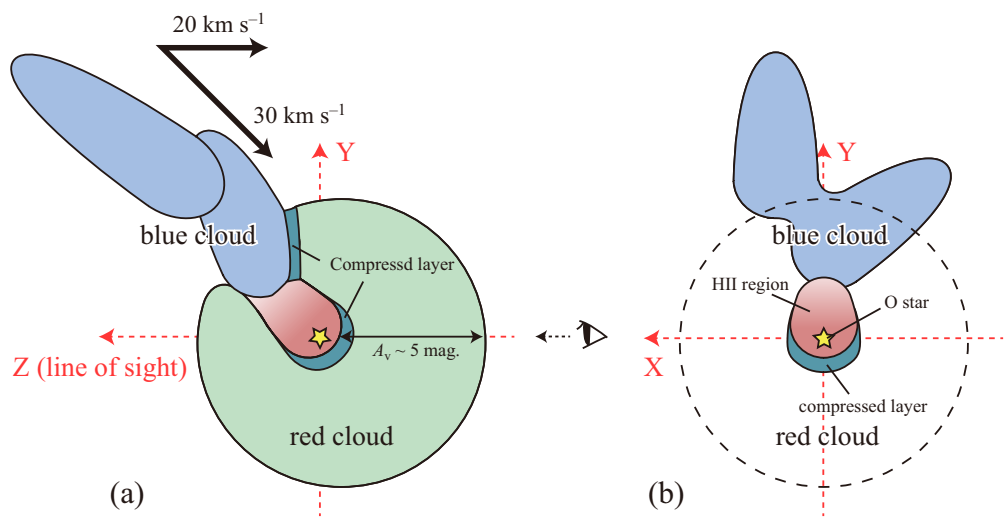}
\caption{Schematics of the evolution of the CCC in RCW\,120 are presented on the Z-Y plane in (a) and the X-Y plane in (b), where the X- and Y-axes are defined as the $-90^\circ$\,--\,$90^\circ$ line and the $0^\circ$\,--\,$180^\circ$ line in Figures\,\ref{radial_plot}(a) and (b), and the Z-axis is along the line-of-sight. The origin of the coordinate system is taken at the exciting O star. \label{ccc_model2}}
\end{figure}

We here discuss an application of the basic CCC scenario to the observed RCW\,120, where the blue cloud corresponds to the small cloud in Figure\,\ref{ccc_model}, and the red cloud to the large cloud.
Figure \ref{ccc_model2} shows an interpretation of the CCC in RCW\,120, where the X- and Y-axes are respectively taken from the $-90^\circ$\,--\,$+90^\circ$ line and the 0$^\circ$\,--\,180$^\circ$ line defined in Figure\,\ref{radial_plot}(a), and the Z-axis is along the line-of-sight.
The origin of the new coordinate system is taken at the exciting O star.
As shown in Figure\,\ref{nasco}(b), the blue cloud has a V-shape structure, each side of which contains one single CO peak at its southern part.
We assume that there was the third peak with a similar size at the south of the present blue cloud and it collided with the red cloud, forming the cavity observed as a ring shape in RCW\,120.
The viewing angle of the collision is assumed to be 45$^\circ$ toward the north from the line-of-sight (= Z-axis).
The colliding velocity is then calculated to be $\sim$30\,km\,s$^{-1}$, and the depth of the cavity is estimated to be roughly 5\,pc with the observed size of the 8\,$\mu$m ring, $\sim$3.5\,pc. 
O star will be formed in the compressed interface layer when sufficient mass is accreted in the high-mass dense cores which are simulated by \citet{ino2013}. 
The interface layer between the two clouds will move at a velocity somewhat less than 30\,km\,s$^{-1}$ due to the deceleration depending on the density ratio between the two clouds \citep[e.g.,][]{tak2014}. 
If we tentatively assume that the decelerated velocity is 15\,km\,s$^{-1}$, the O star will move at the same velocity in the same direction with the small cloud. 
The observed position of the O star with an offset from the center of the ring to the southwest on the symmetry axis of the cavity is then a natural consequence of the triggering under the cloud motion to the southwest. 

\begin{figure}
\epsscale{1.}
\plotone{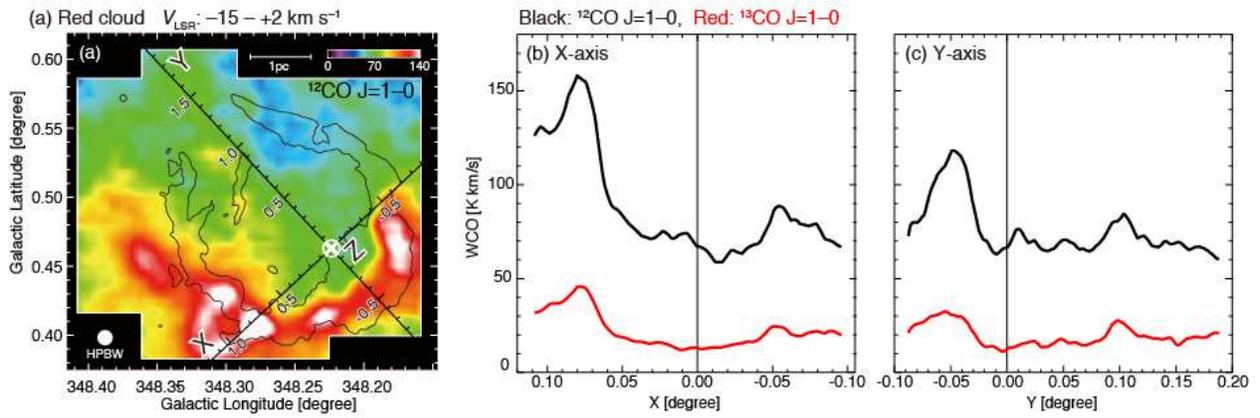}
\caption{CO integrated intensity distributions of the red cloud along the X-axis and the Y-axis are shown in (b) and (c), respectively. The black lines and the red lines respectively indicate the $^{12}$CO and $^{13}$CO. The NANTEN2 $^{12}$CO integrated intensity map is shown in (a), in which the X- and Y- axes are plotted with ticks. \label{slice}}
\end{figure}

After forming the O star in the shocked layer, the UV feedback immediately ionized the surrounding neutral material and prohibited formation of other stars via ionization of the neutral gas, followed by the rapid evolution of the H{\sc ii} region. 
The H{\sc ii} region finally filled out the cavity as observed, indicating that the CCC in RCW\,120 is currently in stage 3 (Figure\,\ref{ccc_model}).
Because much gas still remains toward inside the 8\,$\mu$m ring, the case 3A is here preferable rather than the case 3B. 
In Figure\,\ref{slice}, the CO integrated intensity distributions of the red cloud along the X- and Y-axes are presented. 
The plots show significant CO detection inside the 8$\mu$m ring both in $^{12}$CO and $^{13}$CO, although it is strongly reduced to be 40\,--\,80\,\% in $^{12}$CO and to be 25\,--\,50\,\% in $^{13}$CO from the ring component. 
In addition, as discussed in section 3.2, the O star is likely on the far side of the red cloud, which offers another support for the case 3A.

Differently from the basic model in Figure\,\ref{ccc_model}, in which at stage 3 the ionized gas seems to easily stream out through the opening of the cavity, the opening of the cavity in RCW\,120 looks being blocked by the bottom of the blue cloud (Figure\,\ref{ccc_model2}).
The column density of the blue cloud is by one order of magnitude smaller than the red cloud, making it easier for the ionized gas to stream out through it.
The H$\alpha$ and radio continuum images in Figure\,\ref{comp}, however, do not show a sign of such streams.
Although diffuse H$\alpha$ emission is seen outside the 8\,$\mu$m ring \citep{deh2009}, it is very weak and is extended only by $\sim$1\,--\,1.5\,pc from the 8\,$\mu$m ring even around the opening, not showing evidence for the strong ionized stream. 
It is thus suggested that we are observing the beginning of such event, as discussed in \citet{zav2007}.
The beautiful symmetric ring structure also supports the discussion; as presented in the numerical calculations in \citet{dal2012b, dal2013}, the neutral medium in a mature H{\sc ii} region may show an asymmetric shape depending on its initial density distribution, suggesting that RCW\,120 is in its early evolutional phase.

The plots in Figure\,\ref{slice} are also used to estimate the mass of the molecular gas accumulated by the collision to form the cavity. By measuring the intensity gap between inside and outside the molecular ring in $^{12}$CO, the accumulated molecular mass is roughly estimated to be 1000\,--\,3000\,$M_\odot$.
On the other hand, the accumulated layer could have included the mass of the partner cloud of the collision.
By assuming the H$_2$ column density same as the blue cloud, $4\times10^{21}$\,cm$^{-2}$, it can be estimated as $\sim$1000\,$M_\odot$ with the radius of the 8\,$\mu$m ring of 1.7\,pc, the total mass of the accumulated layer is estimated to be $\sim$2000\,--\,4000\,$M_\odot$. 
This estimate is equivalent to $\sim50$\,--\,95\,\% of the mass of the ring component, $4.2\times10^3$\,$M_\odot$ (see section 3.5), whereas the mass of the ionized gas is estimated as $\sim140$\,$M_\odot$ in section 3.5, which only accounts for less than 10\,\% of the estimated mass of the compressed layer. 
These estimates suggest that ionization by the O star for the compressed layer created by the CCC is minor in mass, and that the remaining huge mass is now observable as the dense molecular ring surrounding the H{\sc ii} region.
This is consistent with our argument in the previous paragraph that the O star and the H{\sc ii} region in RCW\,120 are young, and suggests that the inner structures of the dense ring component were formed before the onset of the interaction with the H{\sc ii} region.

By comparing the cavity depth 5\,pc and the initial collision velocity of 30\,km\,s$^{-1}$, the lower limit of the timescale of the cavity creation and the subsequent O star formation in RCW\,120 can be estimated to be as short as $\sim$0.2\,Myr, which is still consistent with the observational limit for the exciting O star in RCW\,120, as discussed by \citet{mar2010} that any age less than 5\,Myr is possible.
This timescale may be larger like 0.4\,Myr if we adopt 15\,km\,s$^{-1}$ as the average relative velocity because the colliding velocity was probably decelerated during the collision as discussed above.
We will investigate the relationship between the cavity depth and colliding velocity into more detail in the forthcoming paper (Torii et al. 2015b, in prep.).
Considering the argument by \citet{ino2013} that the effective Jeans mass and probably the mass of the forming stars are proportional to the third power of the colliding velocity, it is likely that the exciting O star in RCW\,120 was formed in a short timescale at the beginning of the collision when the colliding velocity is still fast.
The mass accretion rate to from the O star is estimated to be $0.5\times10^{-4}$\,$M_\odot$\,yr$^{-1}$ for a stellar mass of $\sim$20\,$M_\odot$ by assuming that the accreting timescale is $\sim$0.4\,Myr.
Such a high mass accretion rate is consistent with those revealed in the numerical calculations of CCC \citep{ino2013} and the theoretical studies on high-mass star formation \citep[e.g.,][]{tan2003, kru2009, hos2009}.
A similar short timescale of $\sim0.1$\,Myr for the high-mass star formation is observationally derived in the super star clusters NGC\,3603 and Westerlund\,1 by careful optical studies of the young star by HST and VLT \citep{kud2012}.
The short timescale in these clusters was studied with semi-analytical models by \citet{dib2013}, in which a strong increase in the star formation efficiency per unit time is required.

We here summarize the CCC in RCW\,120.
About 0.4\,Myr ago, the collision occurred between the large red cloud with size 12\,pc and mass $5.1\times10^4$\,$M_\odot$ and the small blue cloud with size $\sim$3.5\,pc and mass $\sim$1000\,$M_\odot$, which was accompanied by the V-shape cloud with $7.5\times10^3$\,$M_\odot$. 
By assuming the decelerated colliding velocity of $\sim$15\,km\,s$^{-1}$ and the cloud motion having an angle of 45$^\circ$ to the line-of-sight, the cavity was rapidly created in the red cloud in a short timescale of 0.4\,Myr, forming the compressed layer with mass of 2000\,--\,4000\,$M_\odot$, whose number density increased to $\sim10^5$\,cm$^{-3}$ from $\sim10^4$\,cm$^{-3}$ through the collision.
The cavity creation stopped inside the red cloud at the depth 5\,pc, where the colliding small cloud was dissipated into the compressed layer.
Massive clump formation and subsequent O star formation proceeded along with the cavity creation at a large mass accretion rate up to $10^{-4}$\,$M_\odot$\,yr$^{-1}$. 
After the birth of the O star, an H{\sc ii} region with mass $\sim$200\,$M_\odot$ was immediately created, and filled out the cavity and began interaction with the surrounding dense layer, which is now observed as the molecular ring.
The dense compressed layer was porous, and so the ionized gas partly ($\sim$25\,\%) streamed out though it \citep{zav2007, deh2009, and2014, wal2011, wal2012}. 
The ionized gas just reached the opening of the cavity where the V-shape cloud is currently located, and now it is the very beginning of the erosion by the ionized stream which can be observed as the complementary distribution between the blue cloud and the H{\sc ii} region.
By assuming the velocity of the ionized gas of 10\,km\,s$^{-1}$ \citep{zav2007}, the timescale to fill the cavity with the ionized gas is calculated to be 0.5\,Myr for the cavity's depth 5\,pc, which is consistent with our estimate of the timescale of the CCC in RCW\,120.

The present CCC model in Figure\,\ref{ccc_model2} predicts secondary CCC in RCW\,120.
As shown in Figure\,\ref{nasco}(c), the CO peak in the right side of the V-shape coincides with a local CO peak in the red cloud.
 In addition, broad velocity features which connect the red cloud and the blue cloud are found in the H{\sc i} emission (Figure\,\ref{nasco_lv}).
The bridging feature in velocity may be understood as an intermediate velocity feature expected in an early stage of CCC.
\citet{haw2015} discussed that, by using the data of the numerical calculations of CCC in \citet{tak2014}, the turbulent motion in the compressed layer can be observed in a position-velocity diagram as intermediate velocity features connecting two colliding clouds.
On the other hand, one single broad velocity feature is identified with the CO dataset at the rim of the V-shape cloud toward the direction of condensation 6 with some YSOs (Figure\,\ref{nasco}c).
These observational signatures may indicate the early stage of another collision between the V-shape cloud and the top of the red cloud as shown in Figure\,\ref{ccc_model2}.
If it is correct, the coinciding CO peaks may correspond to the compressed layer by the collision.
Unfortunately, our Mopra and ASTE observations did not cover these coinciding peaks and high velocity features, and in order to investigate the present idea and the possibility of the triggered star formation in future, further detailed observations of a large area are necessary.

Formation of the second generation stars in the context of CCC is another issue to be addressed. 
YSOs and their evolutional stages in RCW\,120 were investigated by \citet{zav2007} based on the {\it Spitzer}/GLIMPSE point source catalog. Subsequently, \citet{deh2009} made identification for the 138 sources whose positions and 24\,$\mu$m brightnesses were iteratively measured from the {\it Spitzer}/MIPS 24\,$\mu$m maps. 
The main difference between these two studies is that in \citet{zav2007} many YSOs are distributed inside the 8\,$\mu$m ring, while in \citet{deh2009} there are almost no YSOs inside the ring. 
Using the carefully measured 24\,$\mu$m stellar samples, \citet{deh2009} applied various criteria for a wide range of wavelength from near-infrared to 24\,$\mu$m, making their YSO identification firm and reliable, and we thus adopt the Deharveng's catalog in the present study.
It is, however, noted that a completeness issue possibly remains in the Deharveng's catalog, particularly toward inside the 8\,$\mu$m ring, because emission from warm dust grains heated by the O star is predominant there, making it difficult to make reliable identification of the YSOs (see Figure \ref{rgb}).
In fact, the statistical studies of the stellar distributions in the Spitzer bubbles indicate significant detection of stars within the bubbles \citep{tho2012,ken2012}.
Analysing their numerical calculations of the triggered star formation by H{\sc ii} region, \citet{wal2013} pointed out that the YSO distributions are peaked around the ionization front, and discussed that the existence of stars inside the bubbles seen in the observational studies can be explained as projection effects of three dimensional structures. 
Same as many other Spitzer bubbles, RCW\,120 possibly harbors the YSOs inside the 8\,$\mu$m ring, although we have to await for future observations to better understand the YSO distribution in RCW\,120.

A most remarkable YSO feature identified in RCW\,120 so far is seen in condensation 1, where a chain of YSOs are aligned parallel to the rim of the 8\,$\mu$m ring \citep{zav2007, deh2009}.
These authors discussed that the YSOs were formed via gravitational instabilities in the collected layer created by a pressure driven H{\sc ii} region.
Other YSOs in RCW\,120 may also be explained by the interactions with the H{\sc ii} region \citep{zav2007, deh2009}.
On the other hand, another possibility is raised in the CCC scenario. 
As already discussed, the high-mass star was likely formed in a short timescale at the beginning of the collision when the colliding velocity was fast.
Then, it is possible that the low-mass star formation may be triggered in a later stage when the colliding velocity becomes even smaller.  
It is interesting to note a speculation that the chain of YSOs located near the symmetry axis of the collision may have been formed under triggering in the final stage of the collision, where the cloud motion was significantly decelerated to a velocity much less than 30\,km\,s$^{-1}$.
The low velocity will lead to smaller turbulent motion and smaller mass accretion rate in the compressed layer, providing a possible alternative to the conventional idea of triggering by the O star ionization.

\section{Summary}
The conclusions of the present work are summarized as follows;

\begin{enumerate}

\item We performed CO $J$=1--0 and $J$=3--2 observations toward the Galactic H{\sc ii} region RCW\,120, which has a beautiful infrared ring, with the NANTEN2, Mopra, and ASTE telescopes, and have identified two molecular clouds associated with RCW\,120 around radial velocities of $-8$\,km\,s$^{-1}$ and $-28$\,km\,s$^{-1}$.
The former (the red cloud) with total molecular mass of $5\times10^4$\,$M_\odot$ has a ring-like component surrounding the H{\sc ii} region in RCW\,120.
The latter (the blue cloud) with total molecular mass of $8\times10^3$\,$M_\odot$ has a V-shape at the north of the opening of the 8\,$\mu$m ring. 
Both of the red cloud and the blue cloud show high $^{12}$CO $J$=3--2/$J$=1--0 ratio of $\gtrsim0.8$ and high kinetic temperatures of $\gtrsim30$\,K, suggesting that they are heated by the exciting O star and/or by the H{\sc ii} region in RCW\,120.
These observational signatures suggest that, despite of a large velocity separation of 20\,km\,s$^{-1}$, the blue cloud as well as the red cloud is physically associated with the H{\sc ii} region in RCW\,120.

\item The large velocity separation cannot be understood as a gravitationally bound system. An expanding H{\sc ii} region is an idea to understand the origin of the molecular gas in RCW\,120. In the assumption of uniform density distribution, Collect \& Collapse (C\&C) is expected to form a dense shell layer, which may explain the observed signature of RCW\,120. 
Our observations, however, indicate no expanding motion in the velocity distribution of the molecular gas, which is inconsistent with what is expected in C\&C. 
The theoretical studies on the non-uniform density distributions also do not provide a plausible explanation for the main observational signatures, i.e., the intensity distributions of the molecular ring in the red cloud, coexistence of the two clouds with a large velocity separation of 20\,km\,s$^{-1}$, and absence of the expanding motion.

\item We postulate the scenario that a collision between the red cloud and the blue cloud triggered the formation of the exciting O star in RCW\,120 in a short timescale of 0.2\,--\,0.4\,Myr. The basic idea of the present cloud-cloud collision (CCC) scenario was first given by \citet{hab1992}.
In this scenario, the velocity separation of 20\,km\,s$^{-1}$ can be interpreted as a projection of the colliding velocity.
Once the collision occurred, a cavity was created in the red cloud, followed by the O star formation in the compressed layer at the interface of the two clouds. The strong UV radiation from the O star then illuminates the inner surface of the cavity, making it observable at 8\,$\mu$m as a ring structure.
The present CCC scenario provides a possible explanation for the present observations, and if this is correct, the RCW\,120 provides the second case of the CCC triggering the formation of one single O star next to M20.

\end{enumerate}

\acknowledgments
We are grateful to the referee to his/her thoughtful comments on the paper.
We are also grateful to Asao Habe for the useful discussion on the cloud-cloud collision.
NANTEN2 is an international collaboration of ten universities, Nagoya University,
Osaka Prefecture University, University of Cologne, University of Bonn, Seoul National
University, University of Chile, University of New South Wales, Macquarie University,
University of Sydney and Zurich Technical University. The work is financially supported by
a Grant-in-Aid for Scientific Research (KAKENHI, No. 21253003, No. 23403001,
No. 22540250, No. 22244014, No. 23740149, No. 22740119, No. 24224005, and No. 23740149)
from MEXT (the Ministry of Education, Culture, Sports, Science and Technology of
Japan) and JSPS (Japan Soxiety for the Promotion of Science) as well as JSPS core-to-core
program (No. 17004). We also acknowledge the support of the Mitsubishi Foundation and
the Sumitomo Foundation. This research was supported by the Grant-in-Aid for Nagoya
University Global COE Program, ``Quest for Fundamental Principles in the Universe:
from Particles to the Solar System and the Cosmos'', from MEXT. Also, the work makes
use of archive data acquired with {\it Spitzer} Space Telescope gained with
Infrared Processing and Analysis Center (IPAC). {\it Spitzer} is controlled by the Jet Propulsion
Laboratory, California Institute of Technology under a contract with NASA.
This research has also made use of data obtained from the SuperCOSMOS Science Archive, prepared and hosted by the Wide Field Astronomy Unit, Institute for Astronomy, University of Edinburgh, which is funded by the UK Science and Technology Facilities Council.

\appendix
\section{CO velocity channel maps of the NANTEN2 and Mopra datasets}
CO $J$=1--0 velocity channel maps of the NANTEN2 and Mopra datasets are presented in Figures\,\ref{channel_all0}\,--\,\ref{channel_all4} at a velocity step of $\sim$2.5\,km\,s$^{-1}$. 
In each figure, the $^{12}$CO $J$=1--0 emission obtained with NANTEN2 and Mopra are shown at the left panels and the middle panels, respectively, while $^{13}$CO and C$^{18}$O J=1--0 obtained with Mopra are presented in the right panels in color and contours, respectively.

\clearpage

\clearpage

\clearpage

\begin{figure}
\epsscale{.9}
\plotone{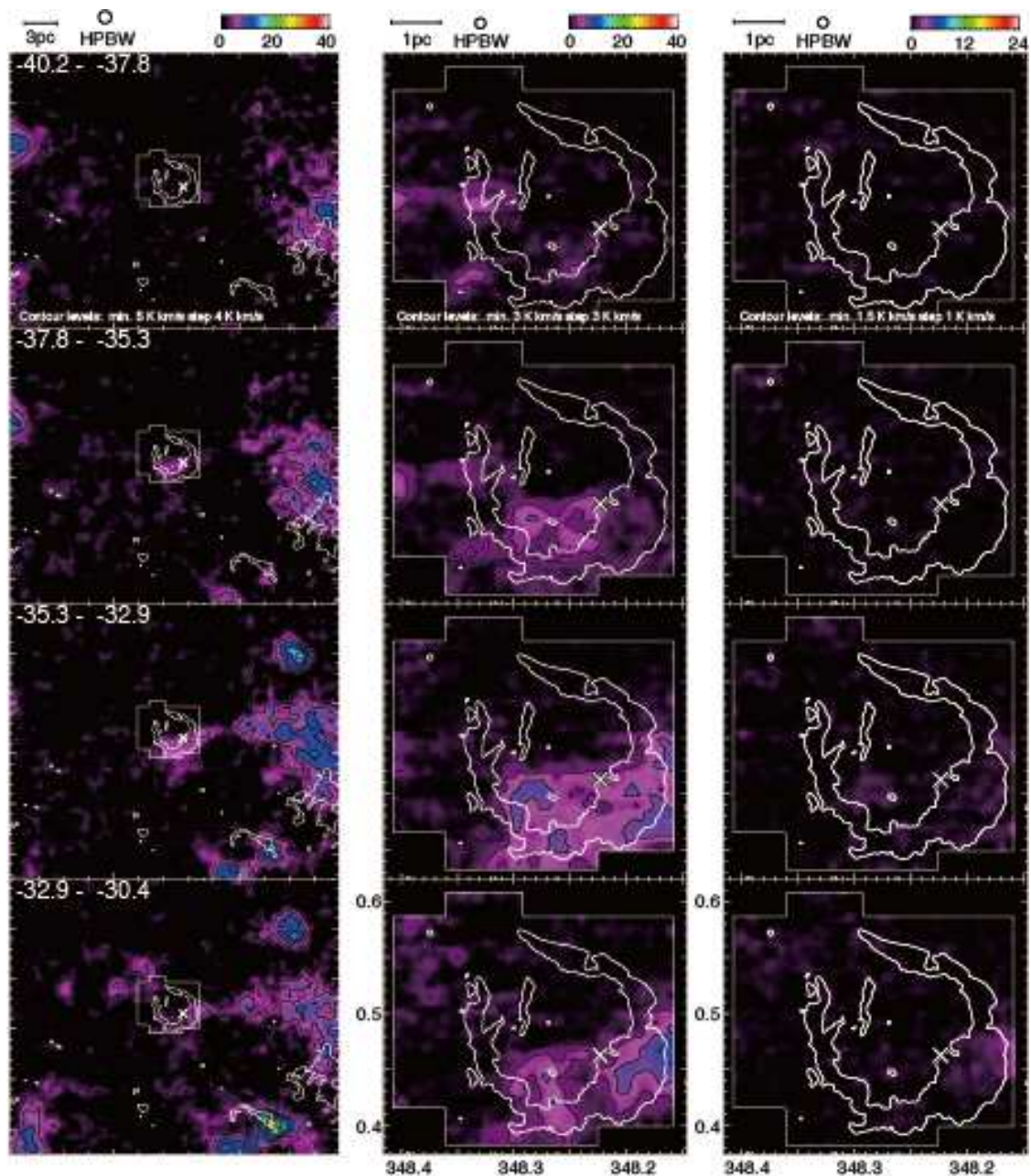}
\caption{Velocity channel maps of the CO emission at a velocity step of $\sim$2.5\,km\,s$^{-1}$.
The central O star is depicted by cross, and approximate boundary of the 8\,$\mu$m ring is shown by white contours.
The Mopra observing area is shown by white lines.
({\it left column}) $^{12}$CO $J$=1--0 (color and black contours) obtained with NANTEN2. 
({\it center column}) $^{12}$CO $J$=1--0 (color and black contours) obtained with Mopra.
({\it right column}) $^{13}$CO $J$=1--0 (color) and C$^{18}$O $J$=1--0 (black contours) obtained with Mopra.
The velocity ranges are shown at the left-top of the left panels in a unit of km\,s$^{-1}$.
 \label{channel_all0}}
\end{figure}

\begin{figure}
\epsscale{.9}
\plotone{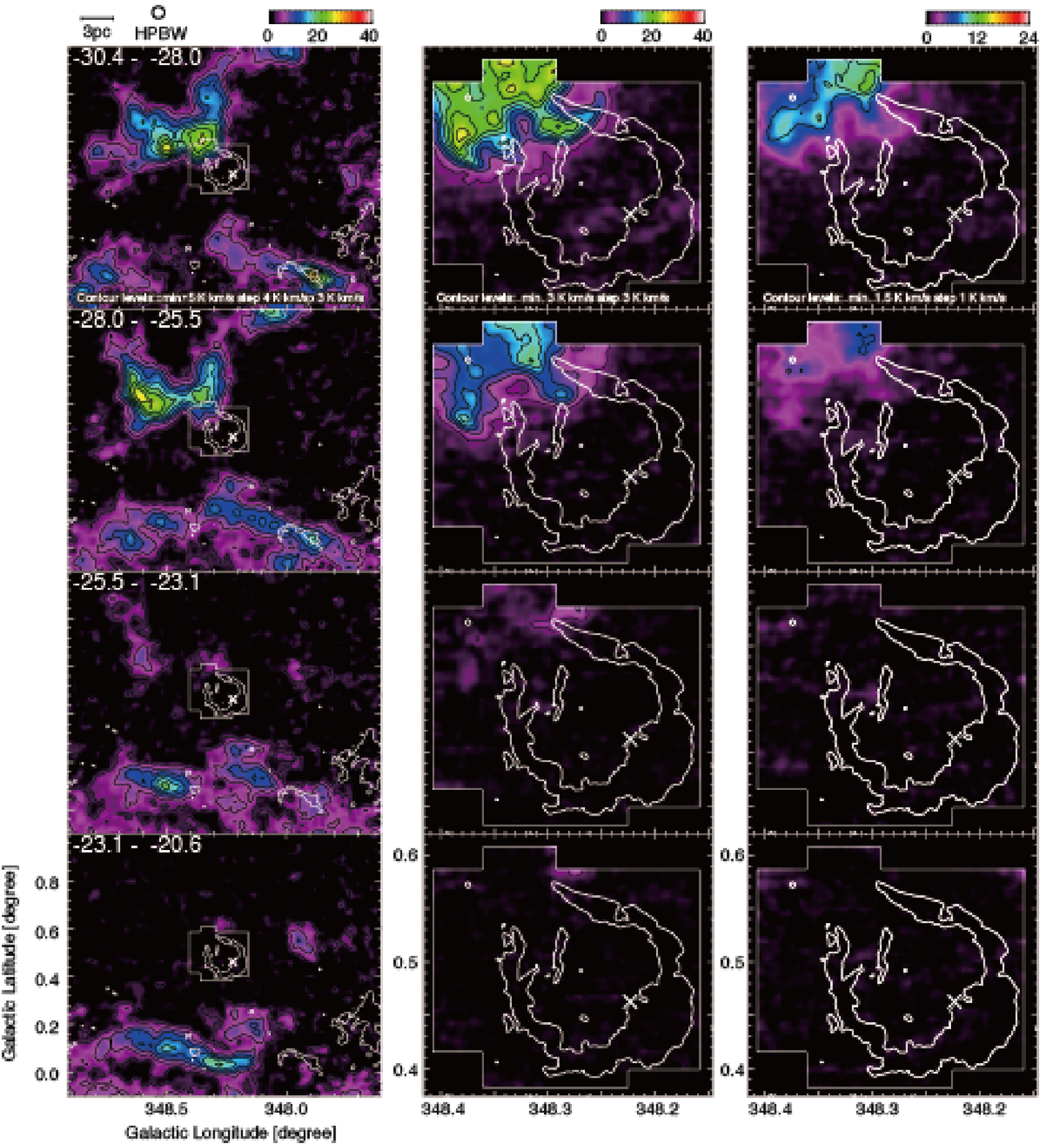}
\caption{Continued.
 \label{channel_all1}}
\end{figure}

\begin{figure}
\epsscale{.9}
\plotone{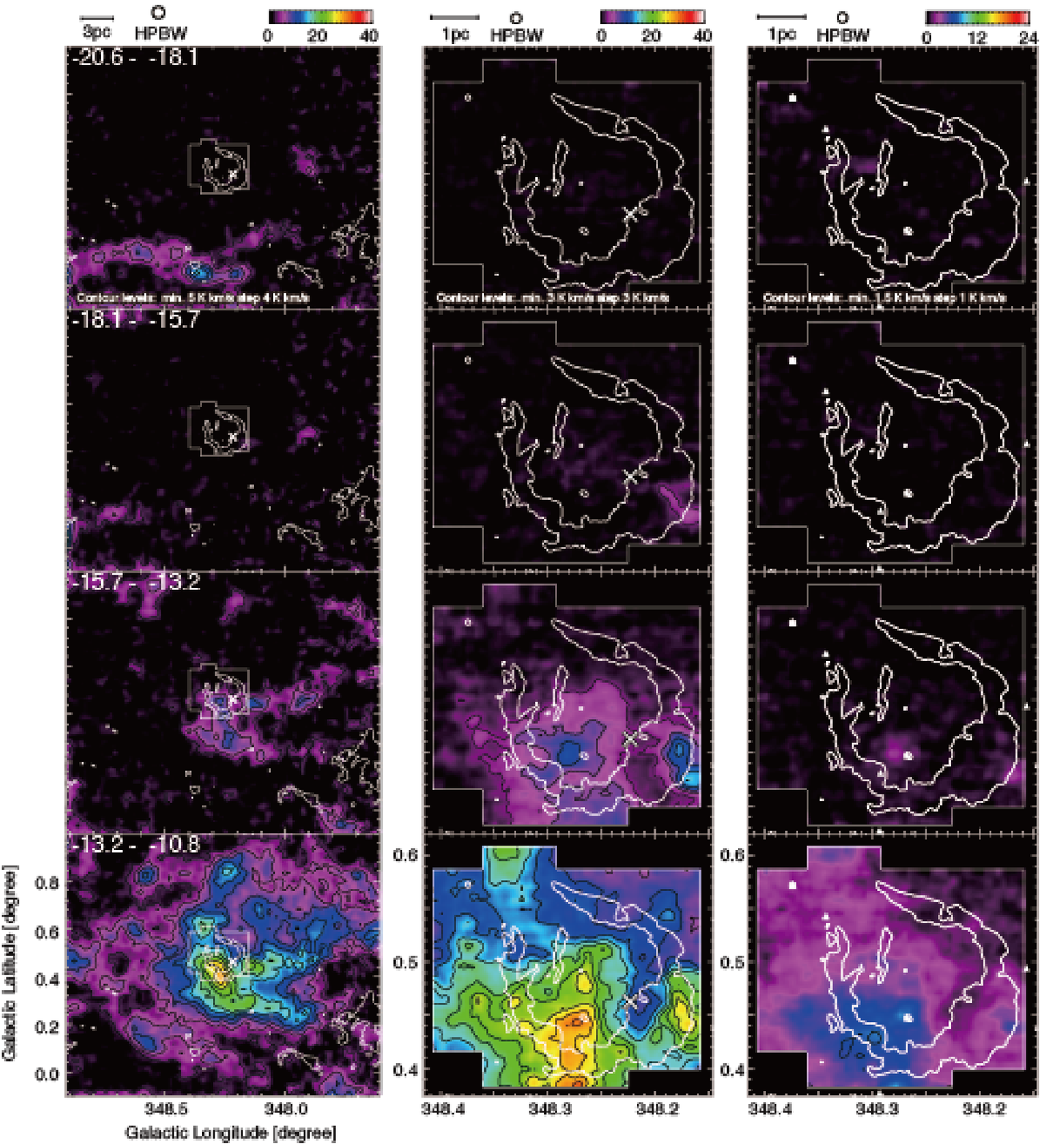}
\caption{Continued.
 \label{channel_all2}}
\end{figure}

\begin{figure}
\epsscale{.9}
\plotone{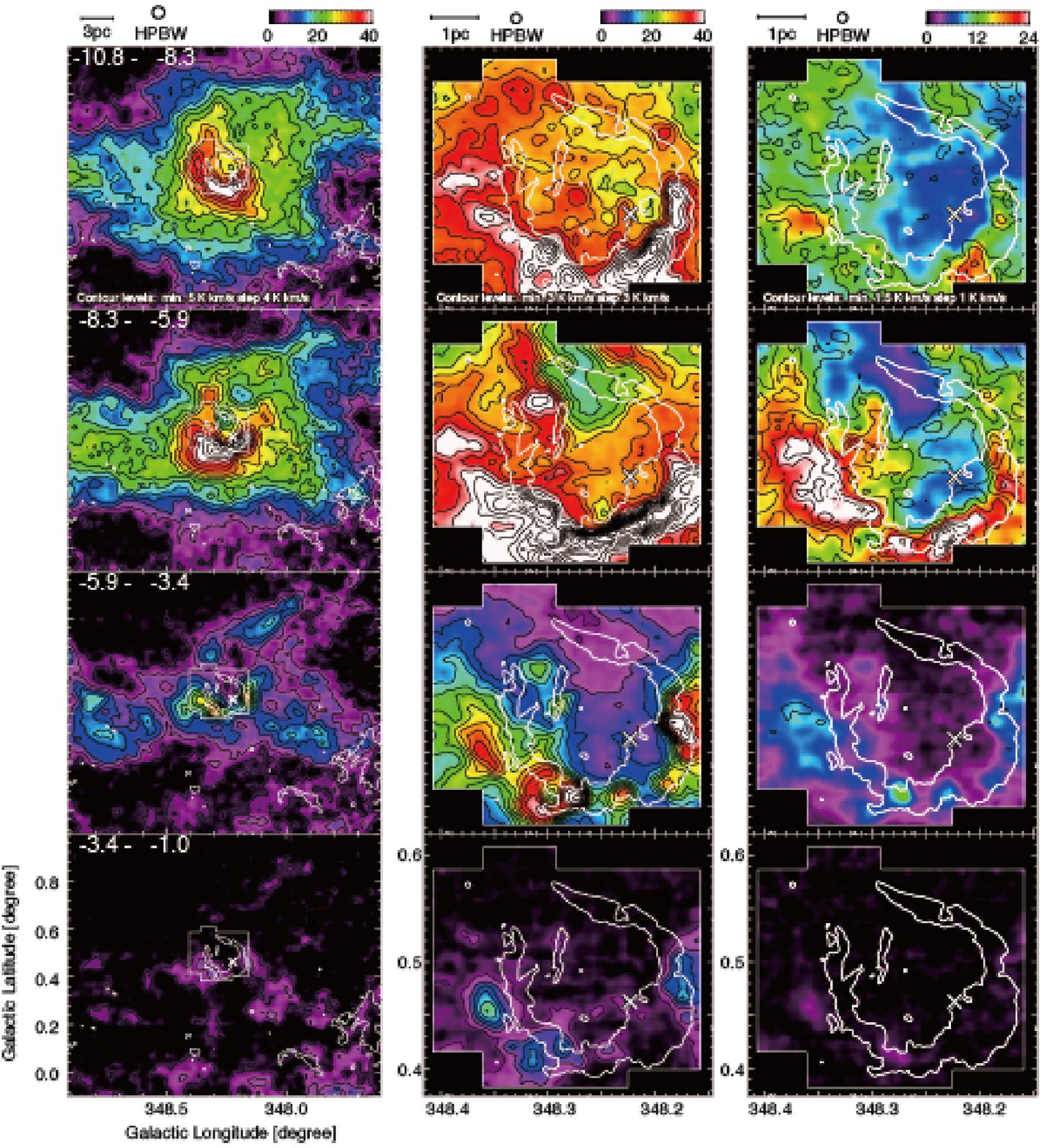}
\caption{Continued.
 \label{channel_all3}}
\end{figure}

\begin{figure}
\epsscale{.9}
\plotone{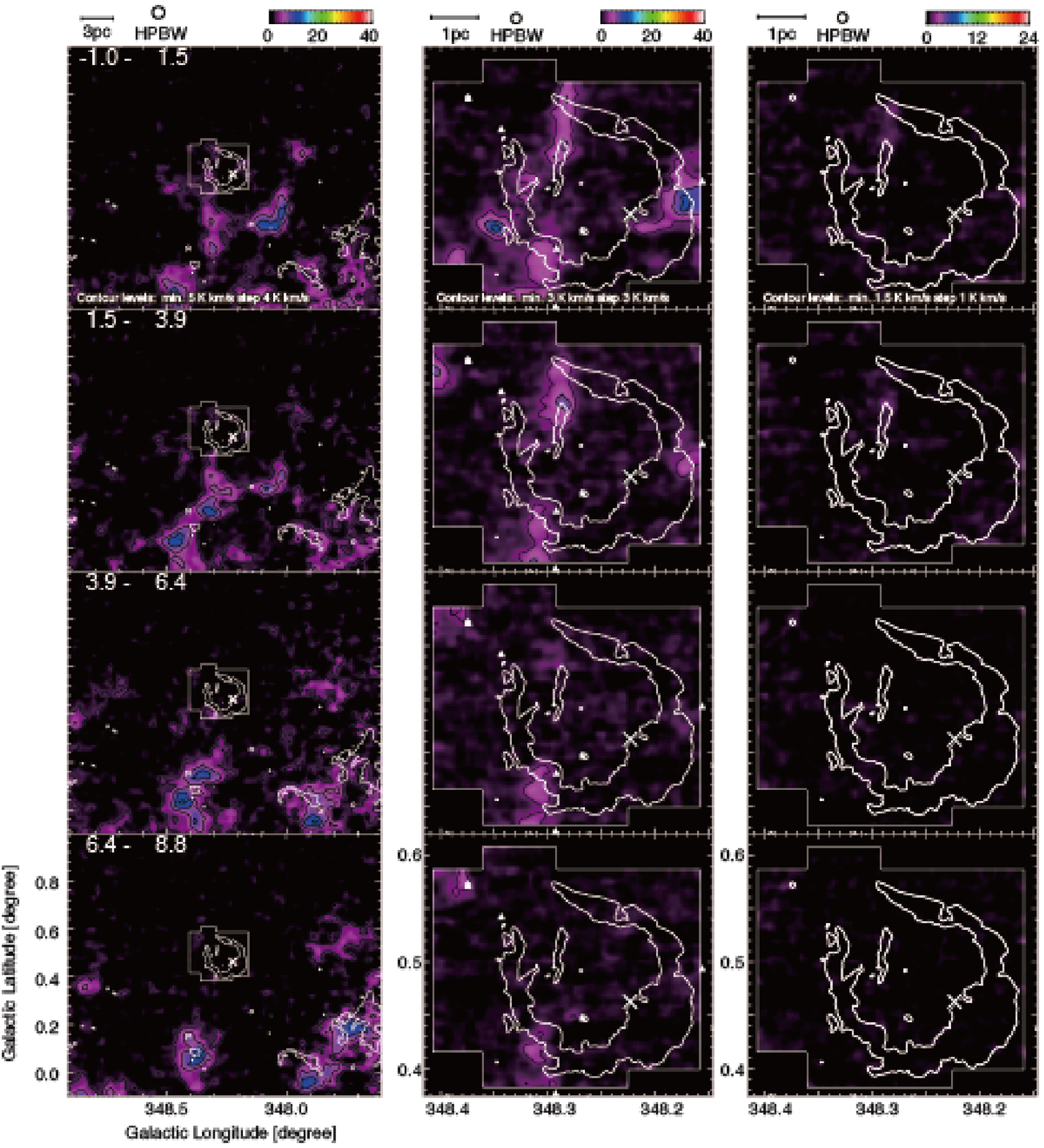}
\caption{Continued.
 \label{channel_all4}}
\end{figure}

\clearpage

\end{document}